\documentclass[10pt,journal,compsoc]{IEEEtran}
\ifCLASSOPTIONcompsoc
  \usepackage[nocompress]{cite}
\else
  \usepackage{cite}
\fi
\ifCLASSINFOpdf
\else
\fi
\usepackage{amsmath}
\usepackage{cite}
\usepackage[hidelinks]{hyperref}
\usepackage{comment}
\usepackage{amsmath,amssymb,amsfonts}
\usepackage{mathtools}
\usepackage{algorithm}
\usepackage{algpseudocode}
\usepackage{textcomp}
\usepackage{xcolor}
\usepackage{cleveref}

\usepackage{amsthm}

\theoremstyle{definition}
\newtheorem{definition}{Definition}[section]
\newtheorem{theorem}{Theorem}[section]

\begin{document}
%
\title{A DLT-based Smart Contract Architecture for Atomic and Scalable Trading}
%
%
%
%

\author{Jan~Kalbantner,~\IEEEmembership{Student~Member,~IEEE,} Konstantinos~Markantonakis, Darren~Hurley-Smith, Carlton~Shepherd, and~Benjamin~Semal
\IEEEcompsocitemizethanks{\IEEEcompsocthanksitem The authors are with Royal Holloway, University of London, Egham TW20 0EX, UK.
E-mail: jan.kalbantner.2018@live.rhul.ac.uk; \{k.markantonakis, darren.hurley-smith, carlton.shepherd\}@rhul.ac.uk; benjamin.semal.2018@live.rhul.ac.uk.}
\thanks{Manuscript created May 6, 2021.}}

\IEEEpubid{This work has been submitted to the IEEE for possible publication. Copyright may be transferred without notice, after which this version may no longer be accessible.}


\IEEEtitleabstractindextext{%
\begin{abstract}
Distributed Ledger Technology (DLT) has an enormous potential but also downsides. One downside of many DLT systems, such as blockchain, is their limited transaction throughput that hinders their adoption in many use cases (e.g., real-time payments). State channels have emerged as a potential solution to enhance throughput by allowing transactions to process off-chain. While current proposals can increase scalability, they require high collateral and lack support for dynamic systems that require asynchronous state transitions. Additionally, the latency of channel initialisations can cause issues especially if fast interactions are required. In this paper, we propose an atomic, scalable and privacy-preserving protocol that enables secure and dynamic updates. We develop a smart contract-based Credit-Note System (CNS) that allows participants to lock funds before a state channel initialisation, which enhances flexibility and efficiency. We formalise our model using the Universal Composability (UC) framework and demonstrate that it achieves the stated design goals of privacy, scalability, and atomicity. Moreover, we implement a dispute process in the state channel to counter availability attacks. Finally, we analyse the protocol in the context of an asynchronous smart grid-based marketplace.
\end{abstract}

}

\maketitle

\IEEEdisplaynontitleabstractindextext

%
\IEEEpeerreviewmaketitle

\IEEEraisesectionheading{\section{Introduction}\label{sec:introduction}}

\IEEEPARstart{I}{n} 2008, Satoshi Nakamoto \cite{nakamotoBitcoinPeertoPeerElectronic2008} presented with Bitcoin the blockchain and therefore the first DLT; a technology with the goal to decentralise the financial system. Since then, many other DLT systems emerged. 

One wide-known application of DLT is cryptocurrency, which gained worldwide attention due to its exponential growth in total market capitalisation \cite{tradingviewinc.CryptoMarketCap2021}. While Bitcoin represents the first generation of DLT systems, newer systems have gained in importance as their functionality allows for more general applications. For instance, Ethereum introduced the concept of Smart Contracts (SC) which allowed the execution of a digital contract without the oversight of a third party. This enables, to name a few examples, the possibility to play publicly available and fair gambling games without any dealer, to have an auction without any auctioneers, or even to have a fair and immutable electoral system without any tallying authority \cite{wangOverviewSmartContract2018}. 

However, the current state of DLT is highly experimental and faces limitations that hinder the wide adoption of these systems. New approaches are required to leverage the resilience, integrity and non-repudiation of DLT without incurring significant ecological and economic costs. 
For instance, the Proof-of-Work consensus algorithm used in many first and second generation DLT systems (e.g., Bitcoin and Ethereum) can have a negative impact on the environment \cite{devriesBitcoinBoomWhat2021}. The increased energy consumption is also one of the reasons many new proposals use less energy-consuming consensus mechanisms such as Proof-of-Stake \cite{stollCarbonFootprintBitcoin2019,salehBlockchainWasteProofofStake2021}.

Furthermore, the permissionless nature of some DLT systems (e.g., Bitcoin, Ethereum) limits the transaction processing speed to a few transactions per second \cite{zhengOverviewBlockchainTechnology2017}. In the case of Bitcoin it is about seven transactions \cite{poonBitcoinLightningNetwork2016} and for Ethereum it is about 15 transactions per second \cite{etherscan.ioEthereumBlockchainExlorer2021}. Undoubtedly, this can lead to problems especially considering that conventional (centralised) transaction systems, such as the Visa network, can process up to 65,000 transactions per second \cite{visaFactSheet}.

With the scalability being a major hindrance for the wide adoption of DLT systems, other measures to increase the scalability should be considered. Scalability is the capability of a DLT to efficiently increase or decrease resources without compromises \cite{dziembowskiGeneralStateChannel2018,mccorryYouSankMy2020,kannengiesserTradeoffsDistributedLedger2020}.

Three main approaches are being pursued to increase the scalability: sharding, novel DLT systems, and Layer-2 (L2) protocols. 
Through sharding \cite{al-bassamChainspaceShardedSmart2017,hellingsCerberusMinimalisticMultishard2020}, the DLT network can be split into multiple partitions (i.e., shards) which has the advantage that peers only need to validate transactions of interest. Novel DLT systems, on the other hand, might increase the network's throughput by utilising newly developed consensus architectures and mechanisms \cite{sompolinskySecureHighrateTransaction2015, eyalBitcoinngScalableBlockchain2016}. 
Lastly, there are L2 protocols \cite{gudgeonSoKLayerTwoBlockchain2020} which, compared to the previously discussed solutions, do not change the trust assumptions of Layer-1 (L1) and also do not modify or extend the underlying consensus mechanism. L2 protocols, however, do enable off-chain processing of transactions through private communication between peers rather than communicating with the DLT system (e.g., blockchain) directly. This approach can reduce the number of transactions that need to be processed by a DLT network while staying compatible to different DLT systems and architectures \cite{backEnablingBlockchainInnovations2014, mccorryYouSankMy2020, gudgeonSoKLayerTwoBlockchain2020}. 
L2 solutions mainly include two proposals: (i) side-chains and (ii) channels. Side-chains are alternative blockchain networks that run in parallel to the main network. Subsequently, a side-chain also requires validators for their transactions (e.g., miners) to uphold the security of the network.
On the other hand, channels \cite{gaiPermissionedBlockchainEdge2019,mccorryYouSankMy2020} such as state channels or payment channels can be considered as $n$ party consensus protocols where each involved party needs to consent to each new state. During a channel process, a DLT protocol is only involved in the start and the closing of such a channel. Therefore, all transactions are exchanged off-chain and no DLT protocol is directly involved. Accordingly, all parties need to do the management themselves (i.e., storing and updating of authorised states). 
A wide-known application for channels is to accelerate payment networks such as Bitcoin's Lightning Network \cite{poonBitcoinLightningNetwork2016} or Ethereum with the Raiden network \cite{brainbotlabsestablishmentRaidenNetwork2020}. Furthermore, several academic proposals \cite{mccorryYouSankMy2020, kalbantnerP2PEdgeDecentralisedScalable2021, gaiPermissionedBlockchainEdge2019, dziembowskiGeneralStateChannel2018} propose the extension of channels to a broader range of applications as they can provide various advantages such as minimising transaction fees, protecting against full collusion, and increasing the scalability of the network (as transactions are processed off-chain and their limiting factor is only the latency between the parties).

While current state channel proposals can increase the transaction throughput and, therefore, increase the scalability, they require the use of high collateral (i.e., locking of coins on-chain while the state channel's interactions are off-chain). In practice, this issue would allow an adversary to conduct a denial-of-service attack which subsequently would hinder users to access their funds. Moreover, the start of these state channels can cause delays especially if quick and flexible reactions are required (i.e., while trading on a marketplace).
Unfortunately, current proposals do not solve the presented issues for state channels.

At the same time, many DLT applications do not consider Quality-of-Service (QoS) which is crucial for tasks where a pseudo real-time processing is necessary.
Current DLT systems focus on permissionless transactions and are unable to meet the network throughput demands of real-time systems. Permissionless transactions may be viable for applications with predictable conditions (i.e., static state changes with defined transition changes) but other applications require more dynamic systems which can handle highly asynchronous state transitions. For instance, the application of DLT systems in energy grids require such a dynamic system. Energy grids are active, complex systems that are unpredictable in the consumption and production of power. State transitions need to be handled fast and efficiently to not disrupt the critical infrastructure and current DLT systems are not able to handle these scenarios.

In this paper, we propose a solution to address those issues and present our protocol which can provide atomicity, scalability, and privacy to its participants. We propose the utilisation of $n$ party SCs and the execution of each transaction in its own state channels. Furthermore, we propose a Credit-Note System (CNS) that allows protocol participants to lock funds in the SC before the creation of state channels which enhances the flexibility of participants.
We evaluate our proposal in the form of a case study that can be used in smart grid Peer-to-Peer (P2P) marketplaces and demonstrate that our protocol achieves its predefined goals by providing a formal model in the Universal Composability (UC) framework.
Specifically, our contributions in this paper include the following:

\begin{itemize}

\item We propose a novel DLT model which combines SCs, and state channels in order to enhance scalability and privacy.

\item We propose the usage of a CNS to increase the flexibility of participants.

\item A hierarchical SC infrastructure is proposed to protect against adversaries and enable the collection of credit notes.

\item We use the UC framework to specify a formal model and show that the protocol realise its goals.

\item With the formal verification tool Scyther, we evaluate the security of the protocol.
 
\item Finally, we discuss the potential use of the protocol within energy P2P marketplaces (i.e., smart grids).

\end{itemize}

The remainder of the paper is organised as follows: 
Section \ref{sec:related_work} reviews relevant literature, Section \ref{sec:system_model} discusses the protocol model and an overview of the protocol is described in Section \ref{sec:protocol_overview}. Section \ref{sec:protocol_description} shows details of the protocol, followed by an evaluation in Section \ref{sec:evaluation}, an adaption to a use-case is presented in Section \ref{sec:application}, and we conclude the paper with Section \ref{sec:conclusion}.

\section{Related Work}
\label{sec:related_work}

Blockchain \cite{nakamotoBitcoinPeertoPeerElectronic2008} as the most well-known DLT is a globally distributed ledger system that can use different consensus algorithms among untrustworthy network participants~\cite{linSurveyBlockchainSecurity2017, zhengOverviewBlockchainTechnology2017, kalbantnerP2PEdgeDecentralisedScalable2021}. The participants are displayed by distributed nodes that can send, store, and verify data. Furthermore, each node has access to the ledger, which contains a copy of the history of all transactions. The use of a consensus algorithm ensures the transactions validity. Only validated transactions are added as a new block to the ledger as a time-stamped chain of blocks. 
Blockchain systems can take on many forms such as private permissioned as well as public permissioned or permissionless systems. While permissioned blockchain systems are controlled and only accessible by previously authorised entities, a permissionless system is accessible without any restrictions \cite{petersUnderstandingModernBanking2015, schletzHowCanBlockchain2020}. 
In general, public and private blockchain systems are distinguishable between the access to information of the ledger. Private systems restrict their access to authorised participants while public systems do not have any access restrictions. 
Note that small systems (e.g., private permissioned systems) might be prone to attacks (e.g., 51\% attack) due to a fewer number of available nodes \cite{kalbantnerP2PEdgeDecentralisedScalable2021}.

While public-key cryptography can provide security and a consensus algorithm can add immutability that prohibit fraud and double-spending \cite{gaiPermissionedBlockchainEdge2019,mohsinBlockchainAuthenticationNetwork2019}, other aspects, however, remain problematic for blockchain-based models \cite{kalbantnerP2PEdgeDecentralisedScalable2021}. 
Two major problems with the technology are scalability and privacy leakage \cite{linSurveyBlockchainSecurity2017}. 

Scalability issues are present in many blockchain systems which motivated academia as well as industry to generate proposals to increase the throughput and efficiency of the networks. 
In blockchain systems, the validation time scales linear with an increase in transactions.
This increase can be traced back to blockchain's block-sizes which scale with the transactions and therefore cause the ledger nodes to expand \cite{kalbantnerP2PEdgeDecentralisedScalable2021}.
One viable solution to increase the scalability of blockchain and other DLT systems is the use of L2 solutions \cite{mccorryYouSankMy2020}. L2 solutions can increase the scalability through off-chain transaction resolution while also protecting sensitive information, as information would be processed private rather than on the blockchain network \cite{liScalablePrivacyPreservingDesign2019}.
There are various proposals regarding L2 solutions but the majority can be categorised into side-chains or channels \cite{backEnablingBlockchainInnovations2014,mccorryYouSankMy2020, eggerAtomicMultiChannelUpdates2019}. In this paper, we focus on channels which include payment \cite{zhangRobustPayRobustPayment2019,malavoltaAnonymousMultiHopLocks2019,eggerAtomicMultiChannelUpdates2019} as well as state channels \cite{mccorryYouSankMy2020, dziembowskiGeneralStateChannel2018,dziembowskiPerunVirtualPayment2017,mccorryPisaArbitrationOutsourcing2019,millerSpritesStateChannels2019,cromanScalingDecentralizedBlockchains2016}.

The major difference between these two kind of channels is that payment channels utilise payment-channel networks (PCNs) which use dedicated nodes to route payment transactions \cite{malavoltaAnonymousMultiHopLocks2019}. State channels, on the other hand, can execute arbitrary complex smart contracts \cite{dziembowskiGeneralStateChannel2018} which makes them more flexible and less prone to attacks but also less reliable.

Existing proposals to extend PCNs include, for example, Egger et al. \cite{eggerAtomicMultiChannelUpdates2019} who proposed the AMCU protocol in order to enable multi-channel updates and reduce the collateral needed for the initialisation of the PCN. Their proposal aims at extending the restricted functionality of existing payment channel networks (e.g. Lightning Network \cite{poonBitcoinLightningNetwork2016}) and tackle the problem of high collateral which can be used in a \textit{griefing attack} to lock coins of a target and deplete the payment channels of competitors. 
They demonstrated that their protocol achieves atomicity as well as value privacy while minimising locked up collateral which subsequently thwarts the threat of the griefing attack. 
However, their work was developed and shown in the context of PCNs which restricts the possible usage of their protocol in other applications. For instance, any applications that require a dynamic and asynchronous state exchange would require the use of state channels rather than payment channels.

State channels, themselves, are not a new concept and other researchers \cite{dziembowskiGeneralStateChannel2018, mccorryYouSankMy2020} already provided detailed information about them. 

For example, Dziembowski et al. \cite{dziembowskiGeneralStateChannel2018} gives an introduction into state channels and also provides the first full specification for general state channel networks. Through channel virtualisation they were able to reduce the latency and cost of state channel networks. 

Moreover, McCorry et al. \cite{mccorryYouSankMy2020} presented an empirical evaluation of single-application state channels which support $n$ parties and allow the channel to be turned off and continued on-chain. Their state channel proposal is a combination of previous works  \cite{millerSpritesStateChannels2019,mccorryPisaArbitrationOutsourcing2019,dziembowskiPerunVirtualPayment2017,cromanScalingDecentralizedBlockchains2016} which also influenced our state channel construction.

While their works \cite{dziembowskiGeneralStateChannel2018,mccorryYouSankMy2020} present detailed information about state channels, they did not elaborate on the usage of state channels in dynamic scenarios where pseudo real-time updates might be necessary. Previous works on state channels disregarded the optimisation of locked up collateral which might be used by adversaries. 
We want to tackle existing problems and extend the use of SCs for collateral mechanisms while using them as `adaptors` to any existing credit-based system. 

There are many blockchain systems that use collateral but for variety of different purposes \cite{scharDecentralizedFinanceBlockchain2020}. 
One of the more common use-case is the cryptocurrency lending market. For example, Yang et al. \cite{yangNewLoanSystem2019} proposed a lending system based on blockchain technology in which the transaction process can be enforced as well as improve insufficient supervision ability and loan efficiency in a transaction process.

Furthermore, Kim \cite{kimNewCryptoSecuredLending2021} proposed a two-way collateral cryptocurrency lending system which allows a borrower to invest parts of their own collateral by predicting the market movements in both directions to make profits irrespective of whether the price of the collateral increases or decreases. The proposed system ensures that the borrower does not loose but might also gain while providing collateral.
Additionally, Yang et al. \cite{yangDesignImplementationLoan2018} proposed a smart contract-based loan system while digitising traditional lending contracts and deploying them into a blockchain.
On the other hand, Shook et al. \cite{shookSmartContractRefereed2019} proposed a fair-exchange protocol for P2P data storage which does not need identity verification. Additionally, they use collateral mechanisms to be able to punish a dishonest participant who abandons the protocol or cheats.

Our work aims to extend previous works \cite{dziembowskiGeneralStateChannel2018,mccorryYouSankMy2020,eggerAtomicMultiChannelUpdates2019} and create the proposal of a new model that can be used dynamically in highly asynchronous and critical infrastructure (e.g. energy grids).

Furthermore, we propose the usage of SCs to aid the management of the state channels in dynamic environments as well as to increase security and privacy through a new collateral mechanism that optimises needed collateral.

\section{System Model and Problem Formulation}
\label{sec:system_model}

\begin{figure}[htbp]
\centering
\includegraphics[width=\columnwidth]{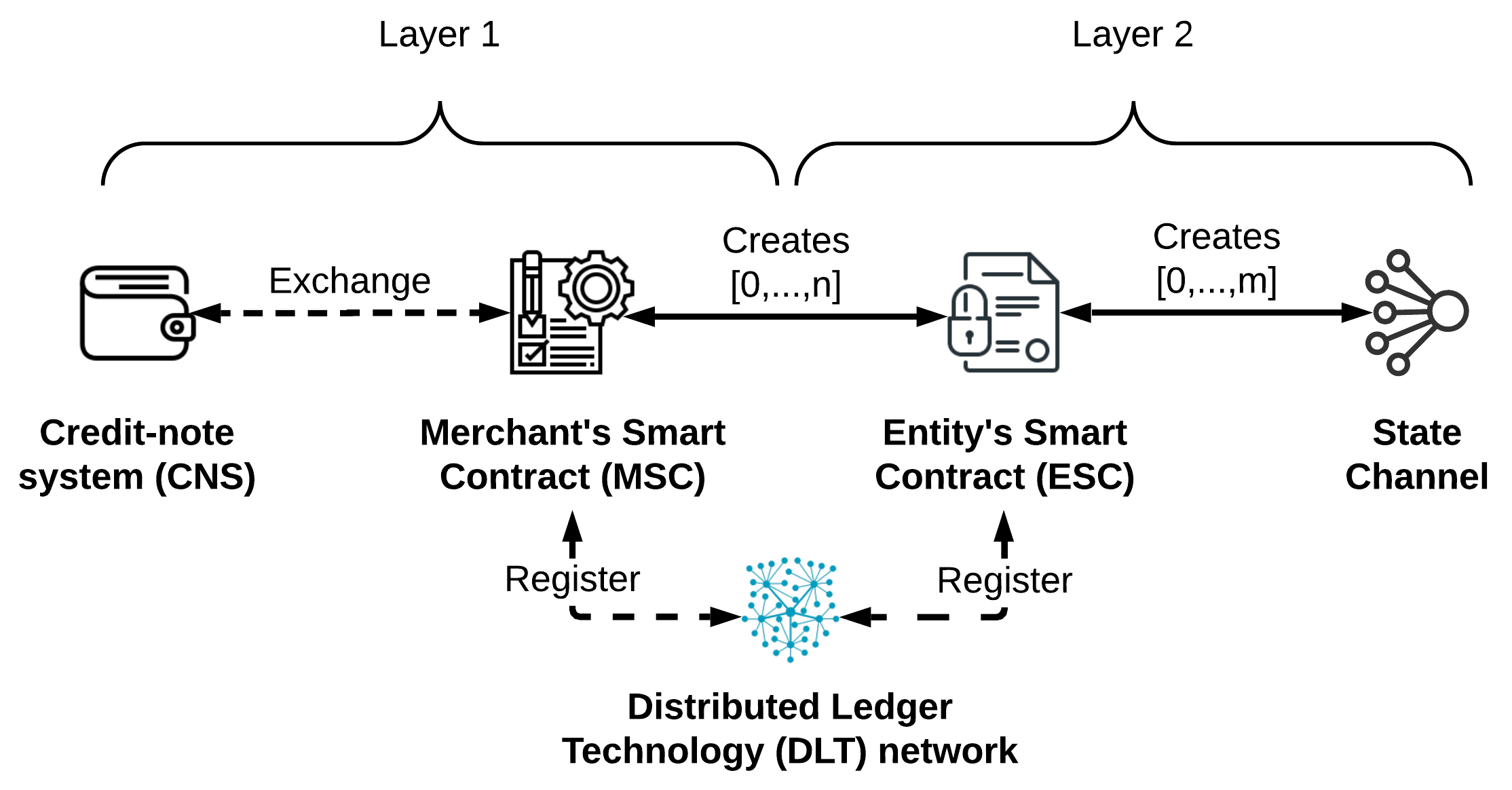}
\caption{Model overview.}
\label{fig:Model_overview}
\end{figure}

For our protocol, we utilise SCs to initialise state channels while protecting locked funds, resolving disputes during an active state channel and updating balances after the channel is closed \cite{kalbantnerP2PEdgeDecentralisedScalable2021}. 
Figure \ref{fig:Model_overview} displays an overview of our model which composes contracts that are arranged in a hierarchical two-layer structure:

\textit{Top-layer}.
    The top-layer is formed by a merchant entity that is connected to a CNS. On the same layer, the merchant creates a SC, which we dubbed the Merchant's SC (MSC). The $MSC$ is a merchant-certified SC which is able to create, manage and execute other SCs that are arranged on a lower layer. Those inherited SCs can be formed between multiple parties and can also include the merchant. 
    
\textit{Bottom-layer}. 
    The bottom-layer is formed by the Entity SC's (ESC's) and the state channels.
    The lower-layer SC's main purpose is to manage and create new state channels between the entities that form the $ESC$. The state channel is used to interact with other parties in off-chain private and authenticated communication channels. 
    The state channels $\mathcal{S} \in \mathbb{S}$ are modelled with respect to a blockchain $\mathbb{B}$ that stores entries as $(\mathcal{P},\alpha^{\mathbb{B}})$ where $\mathcal{P}$ is a user's blockchain address and $\alpha^{\mathbb{B}}$ is the on-chain balance. Furthermore, $t^{\mathbb{B}}$ is the current timestamp in the blockchain and $\mathbb{B}[\mathcal{P}]$ returns the on-chain balance of $\mathcal{P}$.

In order to create state channels, funds must be present during the initialisation of the state channel. Participants in the protocol can decide whether to use the CNS or to lock their own funds for the active state channel phase. If they chose to use the CNS, they need to lock collateral with the merchant in the $MSC$ before any state channel is created. If they do not use the CNS, a previously arranged number of coins (or money) needs to be locked on-chain as a deposit to initialise the state channel. Once this balance is depleted, the result is added as a transaction to the $ESC$ which passes the final sum along to the $MSC$. Similarly, for the CNS mechanism high-resolution credit notes are passed from the closed state channel onto the $ESC$ which transmits them to the $MSC$. These credit notes can then be paid either in a set time interval or locked collateral is used to pay the outstanding debt.

\subsection{Problem Formulation}

Current DLT systems focus on permissionless transactions and, at the same time, they also do not consider Quality-of-Service (QoS) which is crucial for tasks where a pseudo real-time processing is necessary. Subsequently, this creates an inability to meet the network throughput demands of real-time systems.
Permissionless transactions may be viable for applications with predictable conditions (i.e., static state changes with defined transition changes) but other applications require more dynamic systems which can handle highly asynchronous state transitions. For instance, the application of DLT systems in energy grids require such a dynamic system \cite{luoDistributedElectricityTrading2019, zhangEnergyTradingDemand2019}. 
Energy grids as active, complex systems are unpredictable in the consumption and production of power. State transitions need to be handled fast and efficiently to not disrupt the critical infrastructure and current DLT systems are not able to handle these scenarios.

For many years, academia \cite{zhangEnergyTradingDemand2019,liConsortiumBlockchainSecure2017,gaiPermissionedBlockchainEdge2019,luoDistributedElectricityTrading2019, kalbantnerP2PEdgeDecentralisedScalable2021} proposed the usage of DLT systems for smart grids; the smart, decentralised power grids of the future. The decentralisation and P2P infrastructure of DLT could be ideal for smart grids, especially if one takes the many interconnected nodes withing these future grids into consideration.

However, one of the key characteristics of smart grids is the inter-connectivity from energy consumers to energy distributors and energy producers \cite{sianoSurveyEvaluationPotentials2019,andoniBlockchainTechnologyEnergy2019,armbrustViewCloudComputing2010}. 
Compared to the traditional power grid, the smart grid needs to provide better resilience as market participants can exchange bi-directional energy and data flows between each other \cite{zhangEnergyTradingDemand2019}. These bi-directional flows are represented by discrete energy flows and create flexible market conditions. For instance, prosumers (i.e., a consumer with energy producing capabilities) require such conditions to assume either a consumer or a producer state. Prosumers then can either sell their energy back to the market (i.e., the energy producer) or to other localised consumers through a dedicated energy exchange \cite{kalbantnerP2PEdgeDecentralisedScalable2021}.

Now, the bi-directional flows create many transactions that need to be sent from one to another entity. For example, an energy producer switching from producing energy to consuming energy and back will need at least eighth transactions counting in the confirmation of the send messages. 
Although, DLT has many advantages, transactions sent with one of the major blockchain systems can be costly due to high fees, slow due to long confirmation times and consume a lot of energy \cite{etherscan.ioEthereumBlockchainExlorer2021}. 
For example, using Bitcoin where the transaction confirmation takes about ten minutes on average \cite{poonBitcoinLightningNetwork2016} these eight transactions would take about 80 minutes until confirmed. For Ethereum, the throughput is linked to the number of fees paid for a transaction; the more fees are paid, the faster the transactions are verified. However, as we would want to keep the fees as low as possible other methods needs to be used to increase the scalability. 
Moreover, both blockchain systems utilise the Proof-of-Work consensus algorithm which creates enormous amounts of energy waste \cite{devriesBitcoinBoomWhat2021}. 
Bitcoin alone is estimated to consume more than 113 TWh in 2021 which equates to the annual energy consumption of countries like the Netherlands or about half of the consumption of all data centres combined \cite{cambridgecentreforalternativefinanceCambridgeBitcoinElectricity2021, chipolinaHardTruthBitcoin2021}.

\subsection{Protocol Goals}
\label{sec:protocol_goals}

As an addition to the properties of common DLT systems such as non-repudiation and decentralisation \cite{linSurveyBlockchainSecurity2017}, we want to describe the notions of interest to our protocol.

\subsubsection{Privacy}

    A key characteristic of most distributed ledgers is the public availability of the exchanged data. Although most blockchain systems are pseudonymous, this property has advantages but can also create privacy concerns. Especially when blockchain identities are traced back to the owner as all data is publicly available. On public blockchains, for instance, statistical tools and web scraping have been utilised to identify Bitcoin wallet owners.
    When considering legislation, such as EU's General Data Protection Regulation (GDPR), the public availability can create regulatory concerns and need to be solved  \cite{sammanTrendBlockchainPrivacy2016,schwerin2018blockchain}. Without the ability to comply with current regulatory requirements, businesses cannot use these systems for their purposes. 
    
    There are various techniques \cite{schwerin2018blockchain} to provide privacy to the underlying DLT system. The properties of anonymity, fungibility and linkability could minimise the risk of private data exposure \cite{kalbantnerP2PEdgeDecentralisedScalable2021}. Examples of such systems would be zk-SNARK (zero-knowledge succinct non-interactive argument of knowledge) \cite{bitanskyExtractableCollisionResistance2012}, ring signatures \cite{noetherRingConfidentialTransactions2016, sunRingCTCompactAccumulatorBased2017, yuenRingCTBlockchainConfidential2020} and bulletproofs (non-interactive zero-knowledge proofs) \cite{bunzBulletproofsShortProofs2018}. However, the total anonymity of senders, receivers and transactions can hinder these systems to comply with Anti-Money Laundering (AML) regulation \cite{naheemExploringLinksAML2019,poskriakov2020cryptocurrency}.

    For our protocol, the protection of users' privacy and data are of importance. However, in order to comply with current legislation we want to limit the privacy requirements to an additional masking of identities. 
    We say that we captured privacy in our protocol if for every newly created SC or for every state channel change, no one other than the protocol participants themselves can determine their identities and their transaction values.

\subsubsection{Scalability}

    One major challenge of any DLT system is to have sufficient scalability while maintaining security and achieve an acceptable degree of decentralisation. Shifting the focus into either direction usually takes away some degree of fulfilment from the others. 
    
    In our protocol, however, we want to increase the scalability of any underlying DLT system by using state channels as L2 solution. In the context of this paper, scalability will be defined as the ability of a system to grow without compromising its functionality and performance \cite{sigristScalabilityReplicabilitySmart2016}.
    The protocol captures scalability if every newly created state channel is not hindering the creation, dispute or closure of other state channels. Furthermore, the creation of new SCs should not be hindered either.


\subsubsection{Atomicity}
    Atomicity for DLT systems means that the execution of a transaction is considered individually (i.e., non-recursive functions are not able to be re-entered before their return values are committed) \cite{kannengiesserTradeoffsDistributedLedger2020}. However, this also means that transactions can only be executed completely or not at all.

    For our protocol, we assume that atomicity is achieved if every output of the state channel was committed before any other transaction can be submitted. 

\subsubsection{Efficiency}
    The software can be seen as efficient when little to no productivity is lost as a result of an operation. However, for our protocol, we see efficiency as a measure for the other properties defined. This means that our protocol would be efficient if the predefined goals are achieved without compromising security, privacy, scalability or atomicity.

\subsection{Ideal Protocol Functionality}

This section described the ideal world functionality of our protocol.

\subsubsection{Adversary}
Let $\mathcal{A}$ be an adversary which tries to maliciously interact with the functionality $\mathcal{F}$. $\mathcal{A}$ is also modelled as an interactive Turing machine that can interact with a user $\mathcal{P}$ through the interface $attack(\cot)$. This interface accepts as input the identifier of an user $\mathcal{P} = \{ P_1, \dots, P_n \}$ and returns the users' inputs to the attacker $\mathcal{A}$. Furthermore, $\mathcal{A}$ is then able to listen to all communication channels of an user $\mathcal{P}$. 
Additionally, we accept non-adaptive Byzantine corruption (i.e., total corruption) \cite{canettiUniversallyComposableSecurity2001} of any efficient party $\mathcal{P}$ to our model. We only assume efficient attackers who can run an attack in probabilistic polynomial time \cite{eggerAtomicMultiChannelUpdates2019, dziembowskiGeneralStateChannel2018, malavoltaAnonymousMultiHopLocks2019, backEnablingBlockchainInnovations2014, mccorryYouSankMy2020}.

\subsubsection{Message Transmission}

For the communication within the model, we use the predefined UC functions of $\mathcal{F}_{SMT}$ and $\mathcal{F}_{SYN}$ to describe a secure transmission functionality respectively the synchronous communication network \cite{canettiUniversallyComposableSecurity2001}.

Let $\mathcal{F}_{SMT}$ be a secure message transmission functionality \cite{canettiUniversallyComposableSecurity2001}. $\mathcal{F}_{SMT}$ is parameterised by a leakage function $l: \{0,1\}^* \to \{0,1\}^*$ which allows an adversary to capture only the leaked information $l(m)$ and not the plain text $m$. Subsequently, the adversarial is not able to read or change the securely transmitted massage. However, $\mathcal{A}$ can still delay the delivery. 

Moreover, let $\mathcal{F}_{SYN}$ be the functionality of the synchronous network. As described by \cite{canettiUniversallyComposableSecurity2001}, $\mathcal{F}_{SYN}$ displays the synchronous communication network which operates within discrete rounds. Messages sent by a party $\mathcal{P}$ in round $i$ are expected to be delivered in the next round $i + 1$. Additionally, all message send by $\mathcal{P}$ will arrive and will arrive unmodified. However, $\mathcal{A}$ is still able to delay messages but cannot change the order in which they are sent. In the model, we count messages sent with an $i$ and answers to the message with an $i'$.

\subsubsection{Environment}

Let $\mathcal{E}$ be a restricted environment \cite{canettiUniversallyComposableSecurity2001} which is controlling the inputs supplied to the parties running the protocol $\pi$. However, $\mathcal{E}$ is also restricted within the use of external identities. In particular, the set of allowed identities can be determined dynamically depending on the execution of the functions of the environment. 

Furthermore, we assume that environment $\mathcal{E}$ is unable to invoke $\mathcal{F}$ that shares states in any way with protocol $\pi$. This, for instance, applies to any smart contract creation, state channel creation, state channel update or dispute process including any operation between parties where one party's balance is not sufficient. For real-world usage, the participants would refuse the further execution of a state channel without sufficient balance. Additionally, if the state channel is between a party and the merchant, an instant additional credit-note could to the participant to settle the balance at a later point in time \cite{dziembowskiGeneralStateChannel2018}. 

For the model, we also assume that $\mathcal{E}$ will never ask to open a state channel that has already been opened before, close a state channel that was closed, update a state channel that is not open, close a channel in an old state if there is a newer state, update a channel while in a dispute, raise a dispute while one was already raised, or open a channel if there are more than a predefined number of disputes remarked in the smart contract $ESC$. However, a non-malicious entity is expected to return $\bot$ when the environment tries to execute one of the above-mentioned tasks \cite{eggerAtomicMultiChannelUpdates2019, poonBitcoinLightningNetwork2016, malavoltaAnonymousMultiHopLocks2019}.

\subsubsection{Ledger Functionality}

Let $\mathcal{L}$ be the functionality for an ideal ledger. As the ledger functionality $\mathcal{L}$ contains public information which is globally available, we need to consider that the ideal ledger functionality is within a global UC model \cite{canettiUniversallyComposableSecurity2007}. 

\begin{figure}[htbp]
\centering
\includegraphics[width=\columnwidth]{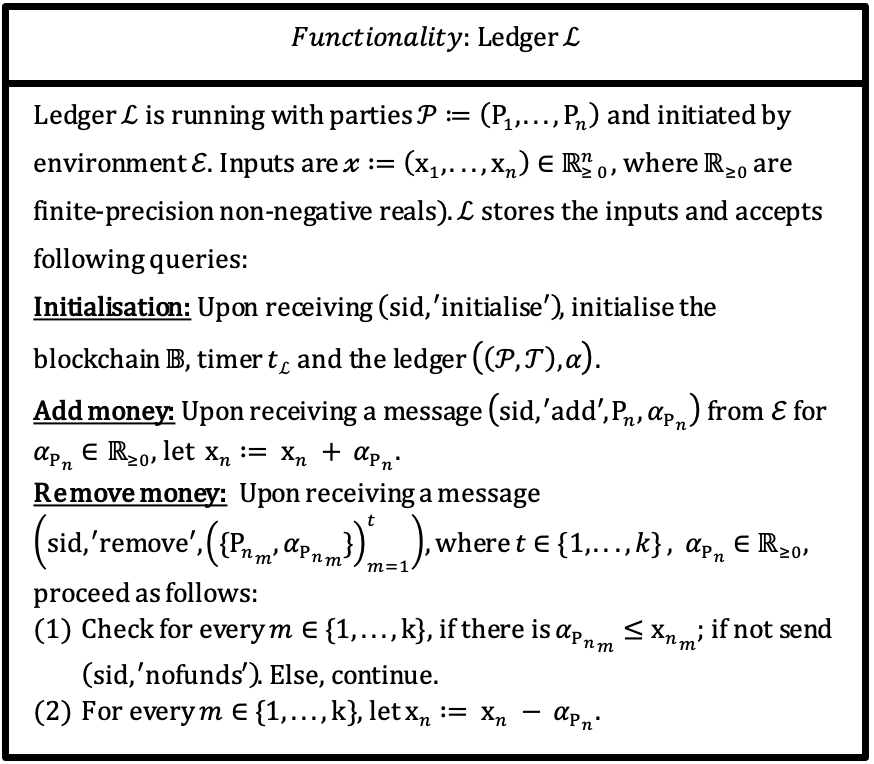}
\vspace*{-5mm}
\caption{Ledger Functionality $\mathcal{L}$.}
\label{fig:func_l}
\end{figure}

As displayed with Figure \ref{fig:func_l}, $\mathcal{L}$ models a DLT system in environment $\mathcal{E}$ which can be used to interact with other ideal functionalities. For simplicity, we will refer to DLT just as blockchain $\mathbb{B}$. Furthermore, the blockchain $\mathbb{B}$ is represented as $((\mathcal{P}, \mathcal{T}), \alpha)$ where $\mathcal{P}$ is a set of addresses that were created at transaction $\mathcal{T}$. Every new session needs a session identifier $sid$ which is short for $sid_i := (sid, i)$.
We define $\alpha$ as the number of coins that $\mathcal{P}$ is holding. Moreover, we have a ledger timer $t_{\mathcal{L}}$ that denotes the current timestamp of the ledger $\mathcal{L}$.

Notice that the ledger functionality might as well be extended to implement a timeout functionality similar to the one proposed by Egger et al. \cite{eggerAtomicMultiChannelUpdates2019}.

\subsubsection{Simulation}

The proposed protocol model in its entirety can display $n$ merchants with $m$ participants which could create as many state channels as their respective computing resources can process. 
For the simulation, however, we operate the model under simplified circumstances to enhance readability and decrease complexity. Let us assume that we only have one merchant $\mathcal{M}$ and two participants $\text{P}_{i}$ and $\text{P}_{j}$ to model the ideal functionality. For the simulation, we do only use the merchant and $\text{P}_{i}$ as $\mathcal{M} \in SC(\mathcal{P})$. Furthermore, we assume that there is only one state channel $\mathcal{CH}$ open at time $t$ between any participants $\text{P}_{i}$ and $\text{P}_{j}$. Additionally, $\alpha_{\text{P}_{i}}$ and $\alpha_{\text{P}_{j}}$ are the deposits of $\text{P}_{i}$ and $\text{P}_{j}$ in a shared channel $\mathcal{C}_{\left\langle \text{P}_{i}, \text{P}_{j}\right\rangle}$. Additionally, let us assume that timeout $t_\Delta$ is agreed upon (e.g., by system timer function). $t_\Delta$ is used to freeze coins during the running of a channel and can also be used to define the deadline for a raised dispute.

\subsubsection{Credit-Note System}
\label{sec:creditnotesystem}

One of the core functionalities of our model (for reference see Figure \ref{fig:Model_overview}) is the credit-note system or CNS which is an optionally-usable system accessible by the participants to have a credit or debit with a merchant.

\begin{figure}[htbp]
\centering
\includegraphics[width=0.9\columnwidth]{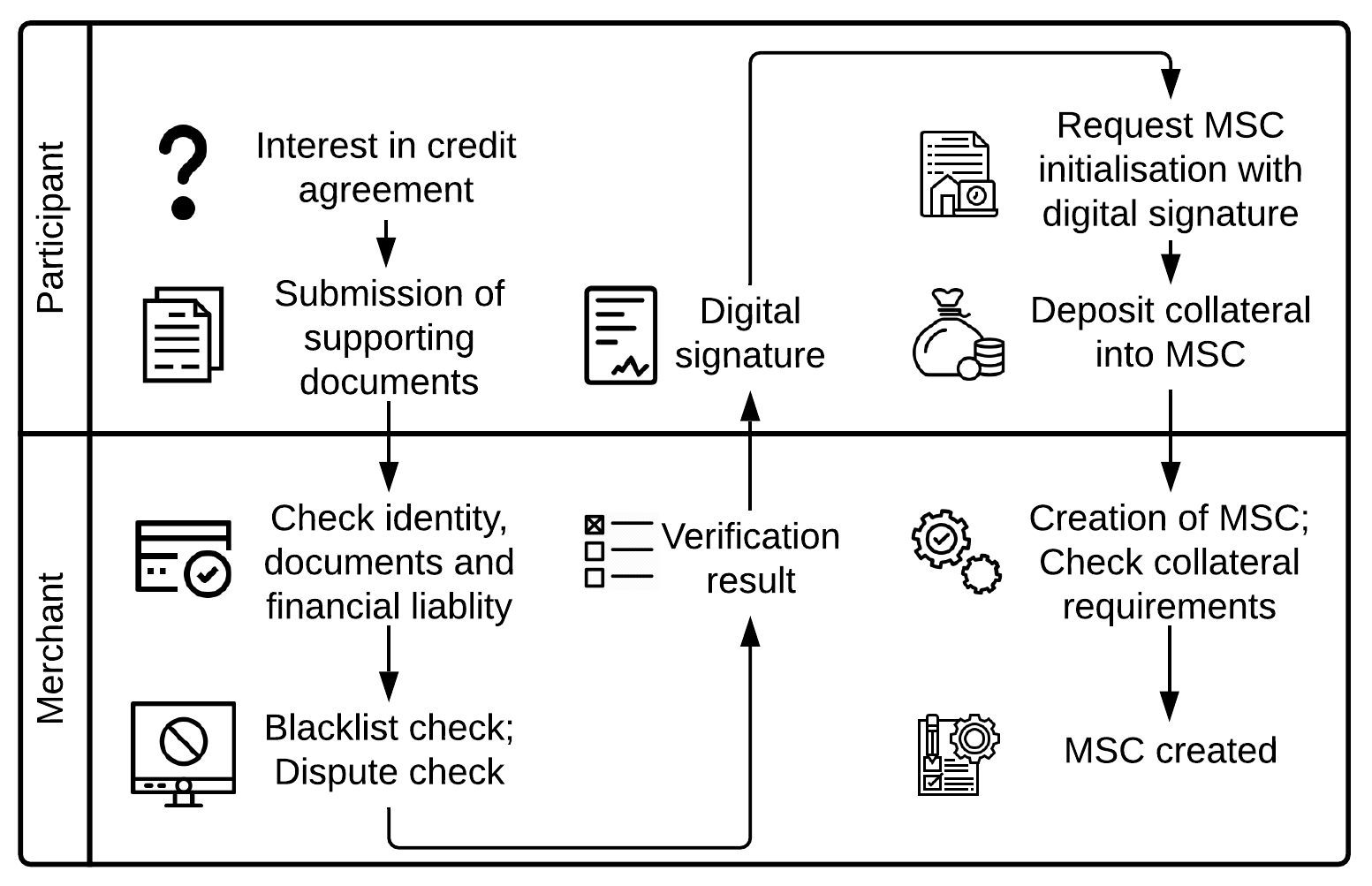}
\caption{Credit-note system (CNS) overview of a successful process.}
\label{fig:CNS_overview}
\end{figure}

Figure \ref{fig:CNS_overview} illustrates the process behind the CNS. Illustrated on the top lane are the participants and on the bottom lane are the merchants. The process begins with participants being interested in a credit agreement. Therefore, a participant has to contact the merchant by sending a request. Enclosed to this request are several supporting documents which are required by law. Note that we will only refer to them as supporting documents as the requirements upon that kind of documentation are country-dependent; however, commonly some sort of identification and bank statements are necessary. Furthermore, this contradicts the privacy requirements from the perspective of the merchant. Therefore, in this case the privacy requirements defined as protocol goals can only be applied to the blockchain transactions.
The received request and the supporting documents are then checked by the merchant. These checks depend on the documents sent but could include checking the identity, checking of bank statements, and most importantly checking for financial liability. This last check is necessary to evaluate the collateral the participant has to submit at a later point.  
After the first step of the verification process, the participant is checked against a blacklist and checked for pre-existing disputes. As the system should only allow participants that are trusted, these kinds of checks are necessary to minimise risks and establish trust. Additionally, some pre-existing cryptocurrency history could be used to verify the participants. When the verification is finished and was successful, a credit agreement can be completed. If the merchant accepts the participant, then a digital signature will be sent which includes the outcome of the verification as well as the needed collateral. This collateral could be the equivalent of a credit score but the threshold for acceptability should depend on the merchant's preferences. 
If the verification was not successful, the participant will still be able to use the protocol but not the CNS with this merchant. Consequently, this unsuccessful participant will either need to try to go through the same verification process with another merchant or lock their funds while initiating a state channel. 
Now, if participants want to create an MSC, they are able to do so by sending the digital signature with the MSC's initialisation message. Furthermore, the participants also need to deposit collateral to the system which they can read out from the digital signature. The collateral amount is defined by the risk of a default the participant poses to the merchant. 
During the initialisation of the MSC, the merchant can then validate the participant based on the digital signature provided.

Notice, that as a collateral either traditional fiat currency or stablecoins (i.e., cryptocurrencies that are pegged to a traditional currency such as USD) should be chosen in order to retain a reliable collateral value. This not only makes it easier to calculate the necessary collateral but also decreases complexity as cryptocurrencies such as Bitcoin are know to have a high volatility \cite{coinmarketcapopcollcCoinMarketCap2021}.
In a best-case scenario, a similar system to the CNS is already existing with the merchant which means also this system could be adjusted to run the proposed protocol.
Nevertheless, there might be privacy issues that arise from the use of static addressing to resolve contract payments which will be addressed in the following sections \cite{naheemExploringLinksAML2019,poskriakov2020cryptocurrency}.

\subsubsection{Functionality}

\begin{figure}[htbp]
\centering
\includegraphics[width=\columnwidth]{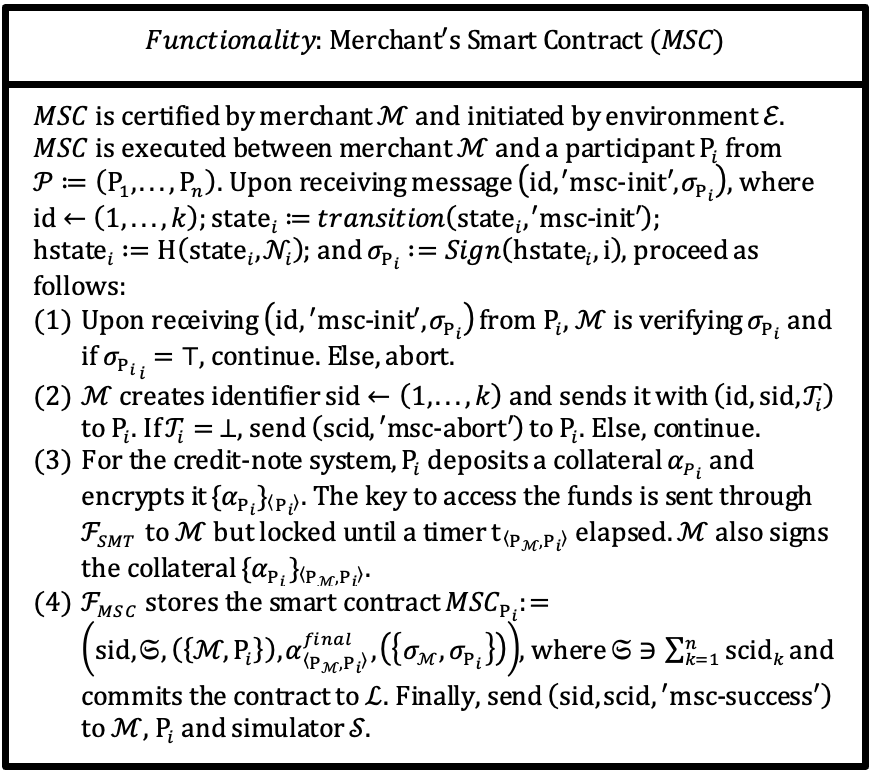}
\vspace*{-5mm}
\caption{Functionality of the Merchant's Smart Contract ($MSC$).}
\label{fig:func_msc}
\end{figure}

\begin{figure}[htbp]
\centering
\includegraphics[width=\columnwidth]{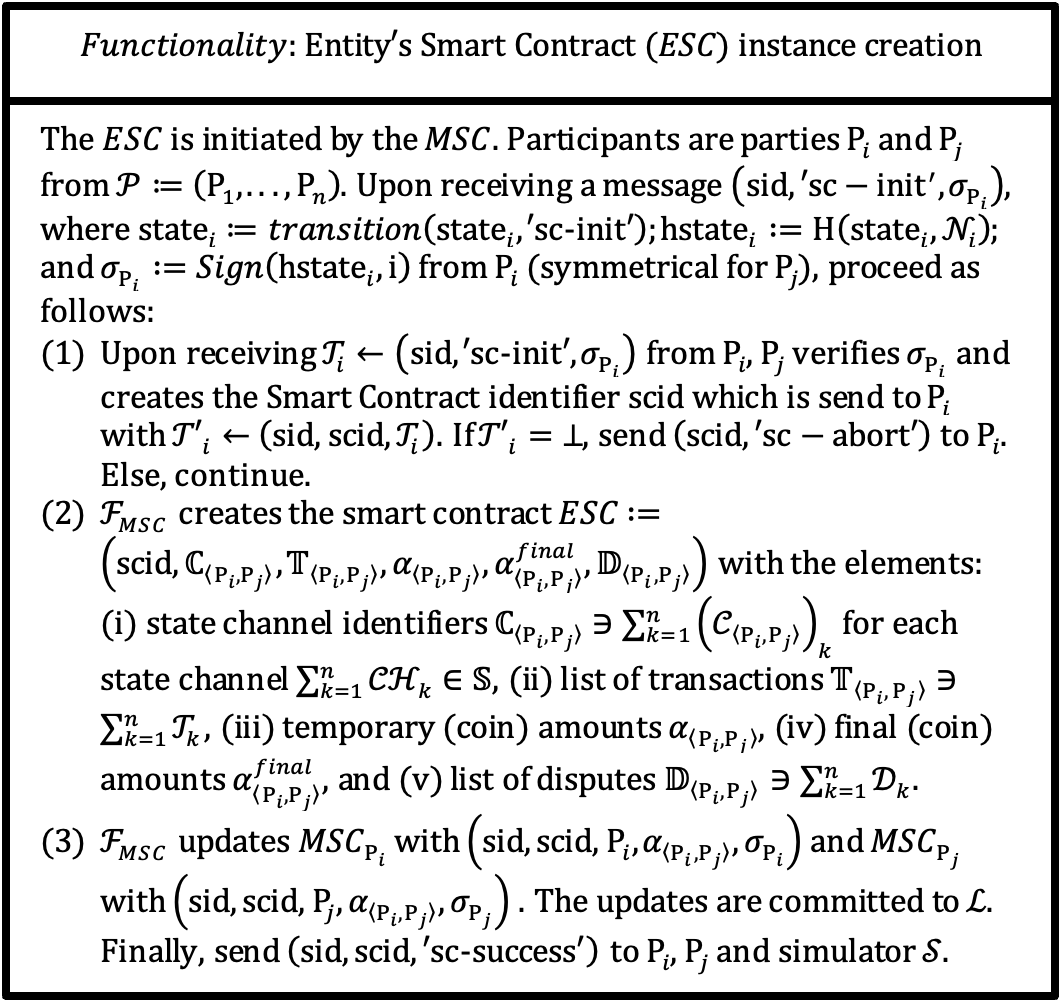}
\vspace*{-5mm}
\caption{Functionality of the Entity Smart Contract ($ESC$).}
\label{fig:func_esc}
\end{figure}

\begin{figure}[htbp]
\centering
\includegraphics[width=\columnwidth]{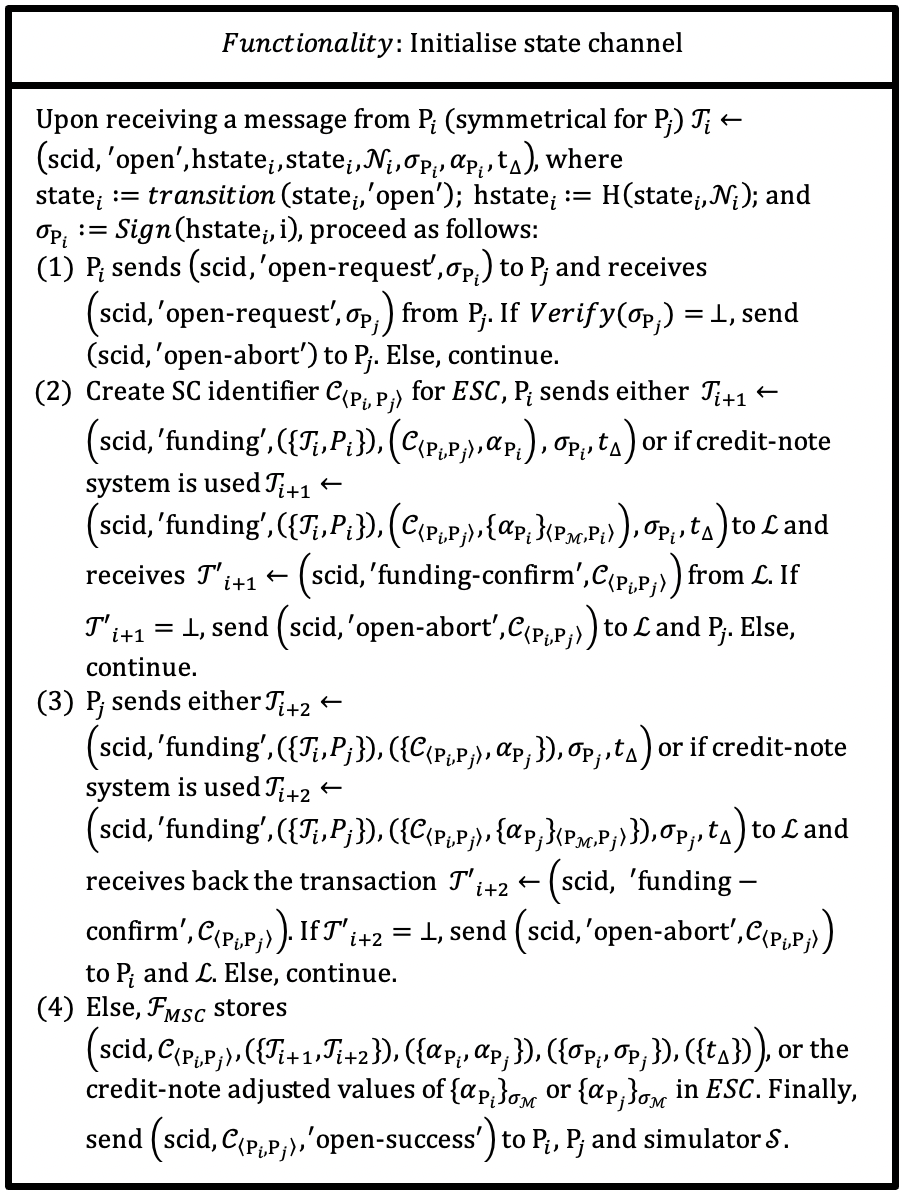}
\vspace*{-5mm}
\caption{Functionality of a state channel initialisation.}
\label{fig:func_init}
\end{figure}

\begin{figure}[htbp]
\centering
\includegraphics[width=\columnwidth]{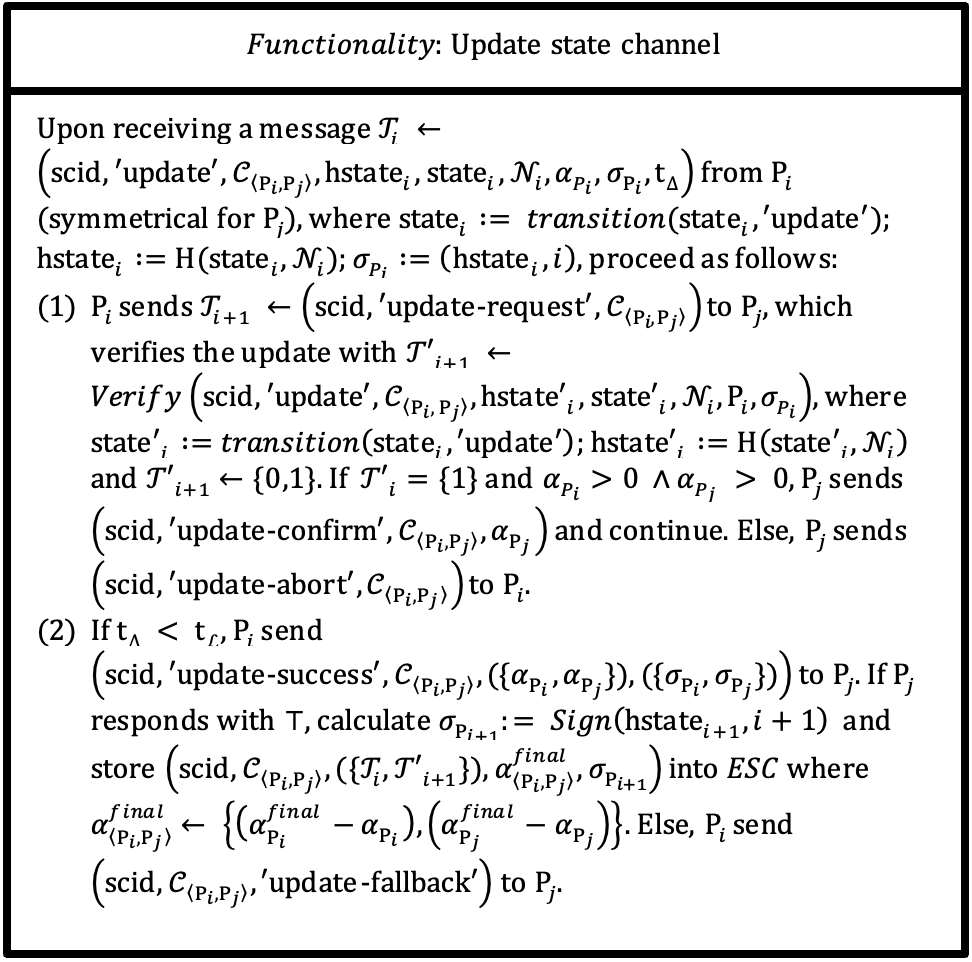}
\vspace*{-5mm}
\caption{Functionality of a state channel update.}
\label{fig:func_update}
\end{figure}

\begin{figure}[htbp]
\centering
\includegraphics[width=\columnwidth]{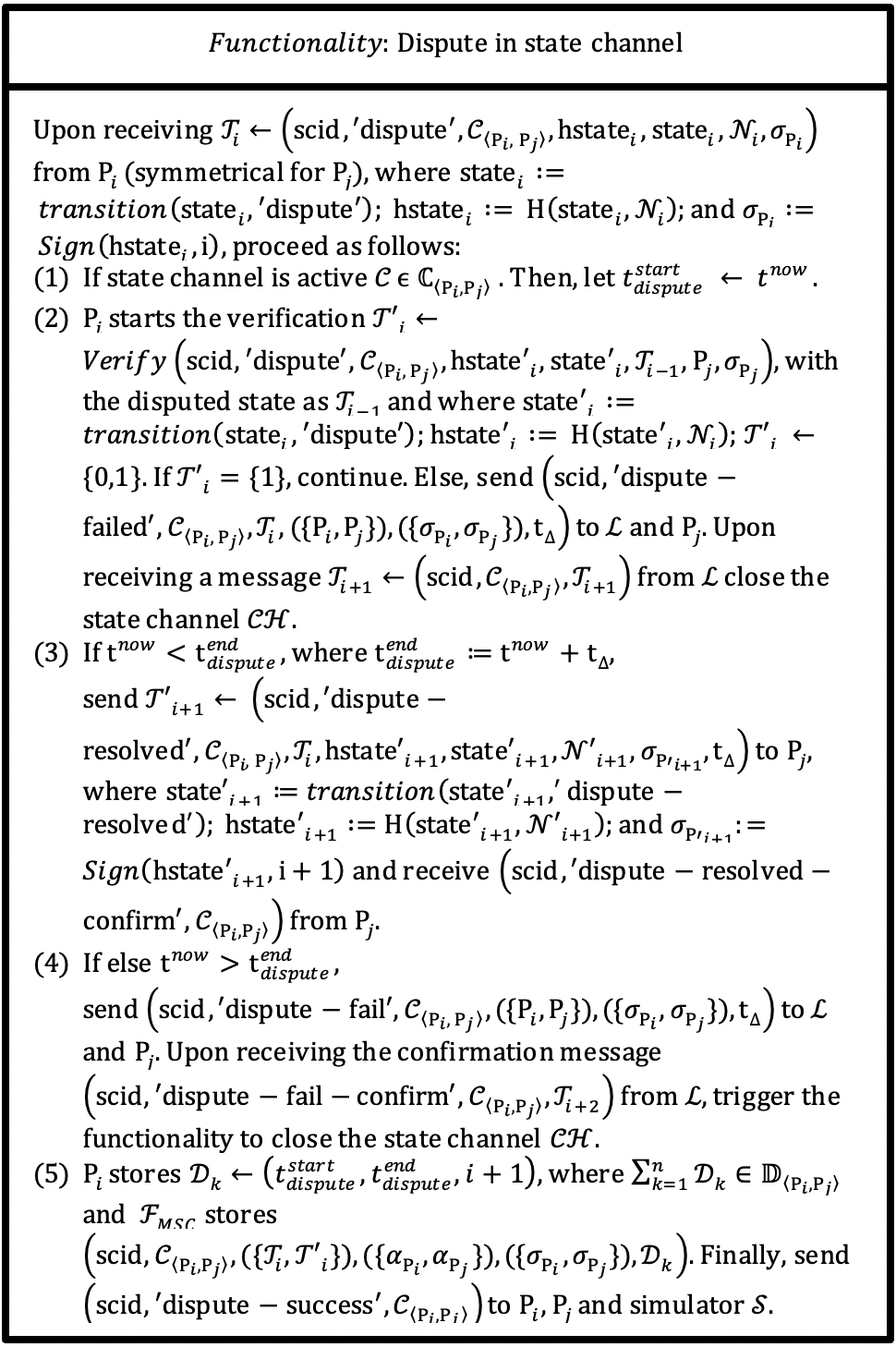}
\vspace*{-5mm}
\caption{Functionality of a dispute in a state channel.}
\label{fig:func_dispute}
\end{figure}

\begin{figure}[htbp]
\centering
\includegraphics[width=\columnwidth]{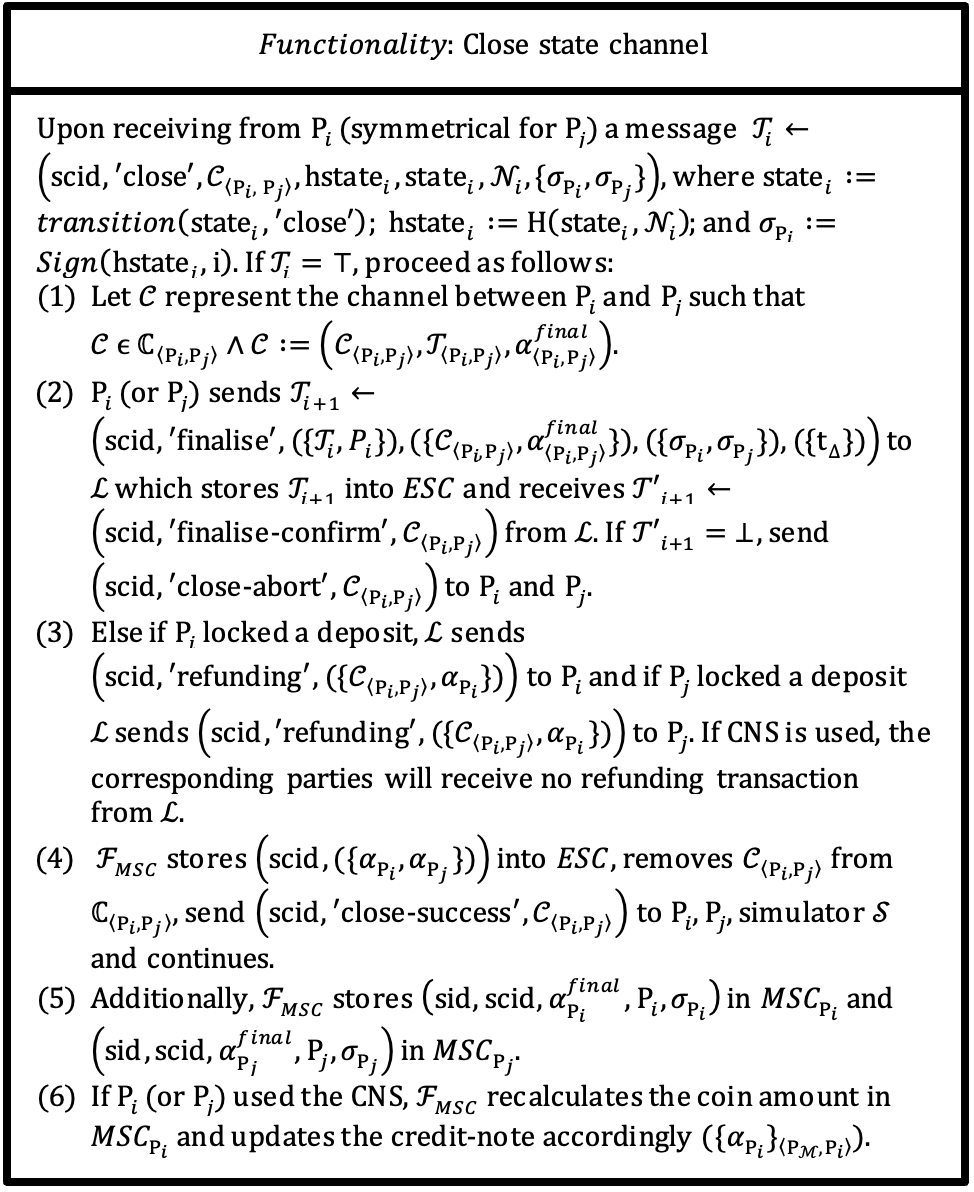}
\vspace*{-5mm}
\caption{Functionality of a state channel closing.}
\label{fig:func_close}
\end{figure}

Let $\mathcal{F}$ be the ideal world functionally which is interacting with users through authenticated and secure channels. The users within $\mathcal{F}$ are displayed through $\mathcal{P}$ and are modelled as interactive Turing machines. 
Each operation of $\mathcal{F}$ is modelled in the UC framework model \cite{canettiUniversallyComposableSecurity2001} and is displayed through Figures \labelcref{fig:func_msc,fig:func_esc,fig:func_init,fig:func_update,fig:func_dispute,fig:func_close}.

In particular, Figure \ref{fig:func_msc} displays the functionality for the $MSC$. The function ensures that a merchant and a participant agree on the creation of a new SC (steps 1 and 2). Furthermore, the participant can use the CNS (described with \ref{sec:creditnotesystem}) and, if decided to do so, must deposit collateral which is further encrypted with a key (step 3). The user then sends the key in a secure channel (e.g. $\mathcal{F}_{SMT}$) to $\mathcal{M}$ which the merchant can use to access the funds after a specified time window has elapsed.
Subsequently, a new $MSC$ is created (step 4) which is used to create other $ESC$s.

Furthermore, the $ESC$'s functionality (described in Figure \ref{fig:func_esc}) is initiated between two participants (step 1). We will describe a use case in Section \ref{sec:application} of how the $ESC$ can be utilised in a P2P marketplace to trade energy contracts. Notice that one of the participants could also be a merchant.
However, a smart contract is created (step 2) and stored in the $MSC$ if the participants came to a consensus (step 3). 

After the creation of individual SCs (i.e., $ESC$s), an operation to initialise a state channel $\mathcal{CH}$ is started (see Figure \ref{fig:func_init}) in which both participants of the $ESC$ agree upon the opening of a new state channel (steps 1 and 2). Afterwards, the funds are locked on the ledger $\mathcal{L}$ or credit-notes are used (step 3) and, finally, a new channel is created (step 4) and information of the state channel are stored in $ESC$. 

Figure \ref{fig:func_update} describes the process of an update in a state channel. In particular, if an update request is sent to a party (step 1), this party then needs to verify the information sent and agree upon the correctness of the proposed state transition. If the previously agreed timer $t_\Delta < t_\mathcal{L}$ (i.e., the timer of the state channel is not elapsed), then the state transitions are executed in the state channels and $ESC$ is updated (step 2). 

Moreover, if a disputed is raised while the state channel is still active, a process is started and run through (see Figure \ref{fig:func_dispute}): A dispute timer is started (step 1), followed by verification operations in which both entities can verify the last state (step 2). If the dispute timer has not passed, the dispute is either resolved (step 3) or not resolved (step 4). In the first case, the details of the resolved dispute are stored (step 5) and in the latter case, the close operation will be triggered. 

Closing a state channel is described in Figure \ref{fig:func_close}. The closing operation is initiated with a verification of the liveliness of a state channel (step 1), followed by the start of a finalisation process (steps 2 and 3) in which all states are aggregated. Additionally, the refunding of locked coins is initiated if CNSs are not used; if CNS are used, no refunding is necessary as the coins are not locked in the state channel. Furthermore, the details of the closed state channel are saved into the $ESC$ (step 4) and also into the $MSC$s of either party (step 5). For the credit-system, additionally, the new coin amount for the corresponding credit-note user has to be calculated, the coin amount updated as well as saved into the $MSC$.

\subsubsection{Universal Composability}

In order to create the formal model of our construction, we use an extended version of the UC framework model \cite{canettiUniversallyComposableSecurity2001} which also includes the ledger $\mathcal{L}$ as shown in \cite{eggerAtomicMultiChannelUpdates2019, dziembowskiGeneralStateChannel2018}.

\begin{definition}
\label{th:definition_uc}
Let $exec_{\pi,\mathcal{A},\mathcal{E}*}$ be the random output of the restricted environment $\mathcal{E}*$ when interacting with an adversary $\mathcal{A}$ and parties running the protocol $\pi$.
Moreover, we consider a protocol $\pi$ that emulates the ideal functionality $\mathcal{F}$ and that is executed by parties from a set of parties $\mathcal{P} = \{ P_1, \dots, P_n \}$. 
The ideal functionality $\mathcal{F}$ is emulated in regards to a ledger $\mathcal{L}$ if for any adversary $\mathcal{A}$ there exists a simulator $\mathcal{S}$ such that for any restricted environment $\mathcal{E}*$, $exec_{\pi,\mathcal{A},\mathcal{E}*} \simeq exec_{\pi,\mathcal{S},\mathcal{E}*}$ (i.e., the distributions are highly similar if not identical to each other). 
\end{definition}

\subsubsection{Discussion}
\label{sec:discussion}

In this section, we discuss the previously defined protocol goals from Section \ref{sec:protocol_goals} and how our protocol fulfils these notions.

\textit{Privacy}.
    Functionality $\mathcal{F}$ only communicates with authenticated channel owners and state channels are only created from an existing $ESC$. Within the $ESC$, participants can use unique identifiers. For example, when trading on a P2P marketplace, parties can use a Universally Unique Temporary Identifier(UUTID) to hide their identities \cite{kalbantnerP2PEdgeDecentralisedScalable2021}. Every party that applies for the CNS with $MSC$, needs to go through a Know Your Customer (KYC) process where the identity, as well as other AML checks, are involved.
    
\textit{Scalability}.
    The main goal of state channels is to increase scalability within DLT systems as most of their transactions are negotiated off-chain. Off-chain transaction processing has the advantage that fewer validations are necessary which therefore increases the scalability. However, scalability and performance also depend on the DLT system which is why the system has to be chosen with consideration. 
    
    Scalability is captured in our protocol if state channels and SCs are not hindering other state channels or SCs, thus slowing down the overall process.
    In order to implement scalability, $\mathcal{F}$ is creating new $ESC$s between new participants. This ensures that state channels can be created independently of any other party and involves only the existing participants. Furthermore, during every step of the $ESC$ creation process $\mathcal{F}$ queries the parties (see Figure \ref{fig:func_esc}). If a participant refuses, the process will be aborted. The same applies to the initialisation of the state channels. In this process, $\mathcal{F}$ also queries the parties before proceeding to the next steps as well as cancelling the state channel if a party refuses to answer. During the finalisation phase, the refunding (if needed) of locked deposits is handled, followed by updating the ledger functionality. Subsequently, the closure can be aborted if something goes wrong. In summary, the protocol is not hindering any participant to create additional SCs or state channels with other participants. Safety measures are installed in case of malicious intent and staling of participants. Notice that there is still the possibility of adversaries to hinder participants from getting their refunds. However, the proposed CNS solves this problem as no coins are locked in state channels. Furthermore, state channels use time locking mechanisms and if participants presume fraud, they can raise a dispute in the existing channel. 
    
\textit{Atomicity}.
    During each step of the protocol (e.g., $MSC$ and $ESC$ initialisation), $\mathcal{F}$ queries each of the parties before proceeding. Therefore, if a participant is not responding or refuses, the step will be aborted.
    However, atomicity is especially crucial for updates and therefore our focus is if the update functionality of the state channel captures atomicity. For instance, in step 1 of the update functionality, $\mathcal{F}$ queries parties before going to the next step. If a party refuses, then the channel will not be updated, every party is notified and the update is aborted. In step 2, $\mathcal{F}$ notifies all parties of the protocol outcome: either successful update or a fallback if the update is unsuccessful (e.g., $t_{\Delta}$ elapsed). Furthermore, the outcome is then either that $\mathcal{F}$ atomically updates the balances of all state channels involved in the protocol or that $\mathcal{F}$ does not update any state channel and notifies the parties. Thus, as $\mathcal{F}$ is trusted, the protocol outcome is the same for all parties, which shows that the model captures atomicity.

\textit{Efficiency}.
    In Section \ref{sec:protocol_goals}, the protocol goals defined that a protocol is efficient if the other properties (privacy, scalability, atomicity) are achieved without any compromise. As described above, the functionality $\mathcal{F}$ successfully achieves the properties of privacy, scalability and atomicity. Thus, we conclude that according to our definition the protocol captures efficiency.

\section{Protocol Overview}
\label{sec:protocol_overview}

\begin{figure}[htbp]
\centering
\includegraphics[width=0.8\columnwidth]{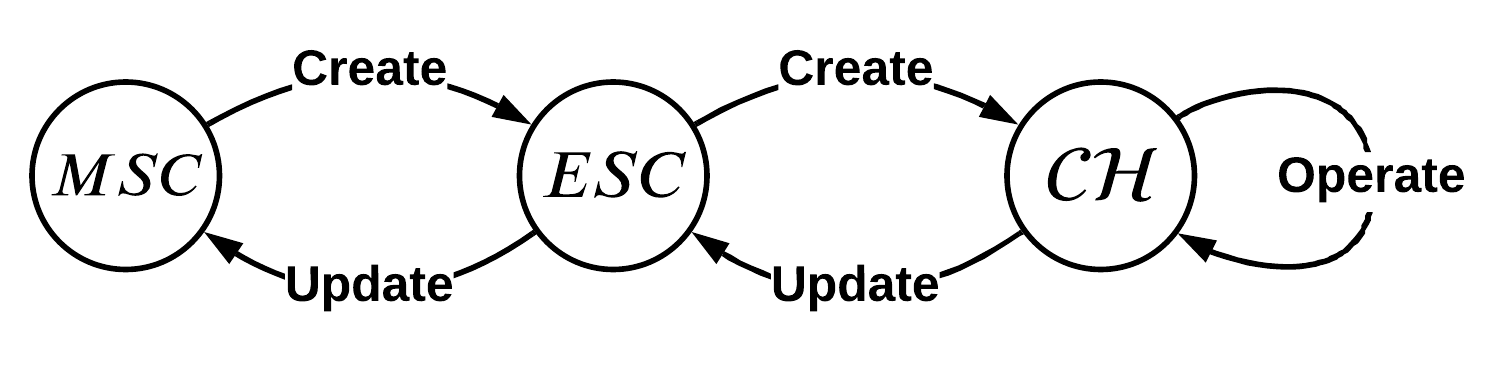}
\caption{Protocol overview.}
\label{fig:protocol_overview}
\end{figure}

The proposed protocol comprises multiple steps where participants can create smart contracts, and can, based on the created contracts, initiate communication in a state channel. An overview of the protocol described is displayed in Figure \ref{fig:protocol_overview}. The steps are structured into the following ones.


1) \textit{$MSC$ initialisation}.
    A party $\mathcal{P}$ wants to participate in a trading process where a merchant is involved. In our example, this could be within a P2P marketplace where the merchant is backed by the CNS. The participants can use the credit-notes which are collateral for the CNS or they can lock their own coins and pay the outstanding amounts directly. 
    However, in the case of the CNS, the merchants would not take the risk of a default by a participant which could create costs for them; thus, a smart contract needs to be made to secure the process for the merchant.
    
    The first step of our protocol is that participant $\text{P}_1$ creates $MSC_1$ with merchant $\mathcal{M}_1$. The smart contract is handled on-chain and therefore could pose a risk to the user's privacy as their details are shared publicly. For privacy preservation, $\text{P}_1$ could use a UUTID and hide the identity when $MSC_1$ is committed to the chain. The consequence, however, would be that the real identity needs to be communicated to the merchant $MSC_1$, for example over an additional off-chain channel. 
    Notice here that the communication of $\text{P}_2$ to $MSC_2$, as seen in Figure \ref{fig:protocol_overview}, is identical to the described process. 
    Furthermore, $\mathcal{M}_1$ is created without any values but with signatures and identities of the merchant and the participant. Additionally, the merchant needs to make sure that any outstanding debt can be collected. Therefore, either a traditional contract needs to be signed or collateral needs to be provided by the participant which can be collected by the merchant. In case the collateral is chosen, a time-lock mechanism should be used which means that the funds are only available to the merchant if a time window $t$ has elapsed. 
    
2) \textit{$ESC$ initialisation}.
    The next step comprises the creation of the $ESC$ (as seen in Figure \ref{fig:protocol_overview}). The $ESC$ is a contract formed between the two parties $\text{P}_1$ and $\text{P}_2$. In this step, the parties create the contract and conduct preparations for the following steps (3), (4), and (5). Part of the preparations is the commit of the deposit to the SC and therefore to the ledger. However, in our protocol, we assume that both participants from the $ESC$ are backed by a $MSC$ which means they are also able to use credit-notes instead of locking a deposit. Therefore, the used credit-notes need then be transmitted to the $MSC$ during the finalisation process (step 5). 
    
    Further, in the protocol, a merchant could also act as a participant which means that an additional $ESC$ could be created between $\text{P}_2$ and $\mathcal{M}_2$ for our example in Figure \ref{fig:protocol_overview}. In that case, $\mathcal{M}$ could provide $\text{P}_2$ directly with a credit as long as there is sufficient collateral for $\text{P}_2$ in $MSC_2$.
        
3) \textit{$\mathcal{CH}$ initialisation}. 
    Step three comprises the initialisation of an off-chain state channel $\mathcal{CH}$. The off-chain capabilities are enhancing the scalability and privacy of the underlying DLT system. 
    As shown in Figure \ref{fig:protocol_overview}, the state channel is created between the two parties $\text{P}_1$ and $\text{P}_2$.
    
    The initialisation step begins when $\text{P}_1$ and $\text{P}_2$ agree upon creating a new transaction $\mathcal{T}_{open}$ which commences the process. Afterwards the funding of the channel by each of the participants through corresponding funding transactions $\mathcal{T}_{funding}$. 
    Any state channel needs collateral which both parties can use while staying off-chain. As described in step 2, both parties can either trade against a locked deposit or can choose to use credit-notes which is added as a negative-sum against the merchant when creating $\mathcal{CH}$. When using credit-notes, the system is still handling the transactions identical as to when a deposit is locked.
    However, only a certain number of coins will be used for the transactions in the state channel. Subsequently, not all available coins from $ESC$ will be locked in a transaction. If additional funds are deemed necessary either a new state channel with the outstanding sum could be created (based on the $ESC$), or the state channel could be closed and a new one created with a greater deposit than the one before. However, for a future paper we want to expand the system and create a dynamic system where an additional credit-note is created when the outstanding sum is too large.
    
    Nevertheless, for the locking mechanism, a previously agreed timer $\text{t}_\Delta$, will be used to lock the deposit until a defined date in future. With the timer, participants can use any previously signed states if an adversary hinders the communication during any future steps.
    Therefore, in our example, the timer is sent with the transaction $\mathcal{T}_{funding}$ so that the locked coins can only be enforced on-chain when the time has elapsed. With sending the signed transactions to the ledger, $\mathcal{CH}$ will become active.
    
    Note that every new state channel will override the current timer which supersedes the last locked state and makes the funds available again. 
    
4) \textit{Active $\mathcal{CH}$}.
    In this step, the participants $\text{P}_1$ and $\text{P}_2$ are trading. While there are still funds available, the participants can exchange goods for coins and vice versa. Accordingly, the update state channel function is used to update their states and later update the SCs ($ESC$ and $MSC$). If, however, credit-notes are used by both participants, then the trading can continue until the loaned amount remarked in the credit-notes is a limiting factor of trading.
    
    Nevertheless, the protocol goals defined in Section \ref{sec:protocol_goals} need to be reached.
    As the initialised channel is communicating completely off-chain, the scalability will already be enhanced because no validator, miner or masternode entities will be necessary to verify any state transition. Additionally, off-chain communication can also increase privacy due to less publicly available information of exchanged transactions. Furthermore, we also consider the property of atomicity which means that either all corresponding coins and goods are transferred or none are.

    During an active state, $\text{P}_1$ or $\text{P}_2$ can create new state transitions by creating transactions $\mathcal{T}_{update}$ which need to be signed by both parties before being valid.
    
    Moreover, during an active state channel session, any party that detects some anomaly can raise a dispute of a transaction by sending a transaction $\mathcal{T}_{dispute}$. A raised dispute can end in one of two ways, either (i) the disputed transaction is verified and therefore the dispute is resolved or (ii) the dispute is not resolved which results in the closing of the state channel.

5) \textit{$\mathcal{CH}$ finalisation}.
    The last phase commences if participants agree upon closing the state channel. Then both parties create a transaction $\mathcal{T}_{finalise}$ (which is also signed by both parties) and send it to the ledger. In the ledger, the $ESC$ and the corresponding $MSC$ are updated accordingly. 
    
    After $\text{t}_\Delta$ elapsed, each state channel has states that can be enforced on the chain which then are accredited to the corresponding parties of the $ESC$. 
    However, if credit-notes were used, the signature of the merchant $\mathcal{M}$ is added to the corresponding transaction. Therefore, $\mathcal{M}$ is the only entity that can enforce the state and will add the state to the corresponding $MSC$ as a debit.

\section{Protocol Description}
\label{sec:protocol_description}

In this section, we formally describe the protocol as well as go into discuss security and privacy.

\subsection{Segments}

This section describes multiple mechanisms that must be present in the used DLT system to run our protocol.

\subsubsection{Time Mechanism}

There are two mechanisms that are used throughout the protocol that utilise time to thwart the activity of malicious adversaries: (i) the time-lock mechanism and (ii) dispute timers.

Firstly, we use time-lock mechanisms throughout the protocol to enforce transactions on the blockchain after a certain time window elapsed (e.g., for the participant's deposit in $MSC$ for the CNS).
In classic blockchain systems, time-lock mechanisms are usually defining time as a block height $h$. This means that if a time-locked transaction is added to the blockchain at block height $h$, then this transaction will be refused by the blockchain miners or validators until $h < h*$ while $h*$ is the current block height. Notice that even though we can estimate the block height, only a rough estimation is possible as blockchain systems are usually non-deterministic biased.

Secondly, we utilise a dispute timer which is used if a dispute is raised in a state channel. This dispute timer is defined with a starting time $t^{start}_{dispute} \gets t^{now}$ and an ending time $t^{end}_{dispute} \gets t^{now} + t_\Delta$. $t^{now}$ defines the current time and $t_\Delta$ is the pre-agreed timer which is used throughout the protocol, for instance with the time-lock mechanism. The dispute timer's purpose is to hinder any adversaries in the state channel from progressing. Instead, the SC closes the state channel automatically if the time frame has elapsed. Furthermore, if the dispute is resolved, the dispute details (including the timer) are written into the state channel and later into the SCs. Disputes that could not be resolved are written into the dispute list of the $ESC$ and the $MSC$ respectively.

\subsubsection{Digital Signature Scheme}
\label{sec:digitalsignaturescheme}

Most DLT systems are based on hash functions and digital signatures as a way to prove ownership of funds and to authorise transfers. Bitcoin and Ethereum, for example, use the secp256k1 algorithm \cite{brownSECRecommendedElliptic2010} with ECDSA to generate their digital signatures \cite{schumacherWallets2019,librehashBIP0340,hotchkissETHEREUMVIRTUALMACHINE2021}.
In order to use DLT systems, our protocol, and especially our off-chain state channel, needs to comply with these requirements by using a digital signature scheme such as secp256k1. 

The digital signature scheme comprises the algorithms ($Gen, Sign, Verify$) which are defined by the following \cite{canettiUniversallyComposableSignatures2003, eggerAtomicMultiChannelUpdates2019}: 

\begin{itemize}
    \item The generator function $key_{s}, key_{v} \gets Gen(1^{\lambda})$ takes the security parameter $1^{\lambda}$ as an input and returns a signing and a verification key $key_{s}, key_{v}$. 
    
    \item The signing function is defined with $\sigma \gets \text{Sign}(\textit{state}, i, key_{s})$ where the inputs are the state $state$, an incremental counter $i$ and the signing key $key_{s}$. Output of the function is the signature $\sigma$. 
    
    \item The verification function is defined by $\{0, 1\} \gets \text{Verify}(\textit{state}, i, \sigma, key_{v})$ with the state $state$, an incremental counter $i$, the signature $\sigma$ and the verification key $key_{v}$ as inputs. As a result the function can return either a 0 for an invalid signature or a 1 if $\sigma$ is a valid signature of $state$ created with signing key $key_{s}$.
\end{itemize}

Please note that the signing and verifying functions are shown simplified as we do not use the state $state$ but the hashed state $hstate$ (i.e., the state with an applied hash function $H(\cdot)$) in the protocol description.

\subsubsection{Disputes}

The protocol proposed captures its goals in terms of security and privacy. However, it not only achieves these goals but also provides the participants with accountability.
In particular, all protocol failures are reported which subsequently allows other parties to react to e.g. inactive parties. Moreover, participants are able to raise disputes which allow to reveal any malicious behaviour and enables parties to hold these accountable. However, parties could also use the dispute functionality to falsely blame other parties, in which case the other party can provide the missing signature and the protocol could continue. 
Additionally, an adversary might try to stall an honest party and force their funds to be locked in the state channel. In order to avoid this case, a dispute timer is used to automatically cancel the state channel if the timer elapses which subsequently would release the locked funds. 

In order to achieve accountability, the disputes are then also written into the contracts. This system could also be extended so that parties can be written into grey or blacklists.

\subsubsection{Billing}

One major advantages in using the CNS is that participants can trade freely and the billing is handled after a pre-defined time-frame. This time-frame is linked to the merchant's CNS and the $MSC$.
The advantage is that the transactions can be collected and after that time-frame elapsed, the participants, as well as the merchant, could be paid either directly or through the collateral. Moreover, the bills will be highly accurate and every transaction can be tracked indefinitely. 
As the participants are registered with the merchant to use the CNS, the merchant will always be able to track down parties and, due to the collateral, payment of every party is ensured. 
If a participant is not using the CNS, the coins will need to be collected immediately.

\subsection{Formal Protocol Description}

\begin{figure}[htbp]
\centering
\includegraphics[width=\columnwidth]{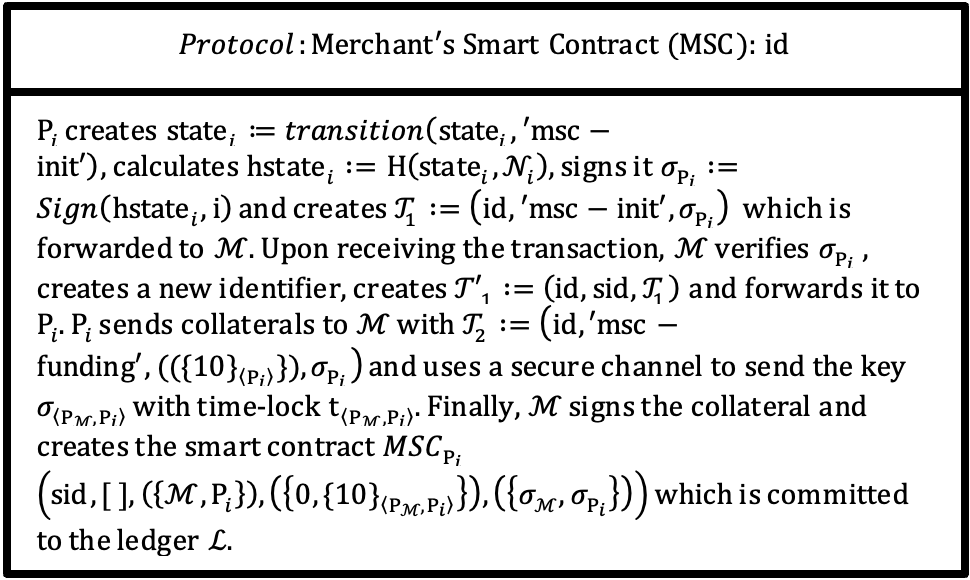}
\vspace*{-5mm}
\caption{Protocol function of the Merchant's Smart Contract ($MSC$).}
\label{fig:prot_msc}
\end{figure}

\begin{figure}[htbp]
\centering
\includegraphics[width=\columnwidth]{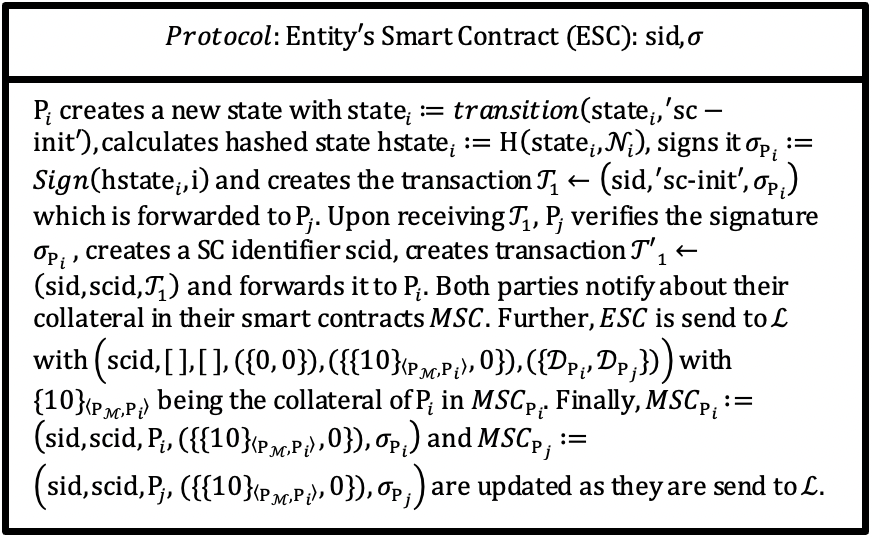}
\vspace*{-5mm}
\caption{Protocol function of the Entity Smart Contract ($ESC$).}
\label{fig:prot_esc}
\end{figure}

\begin{figure}[htbp]
\centering
\includegraphics[width=\columnwidth]{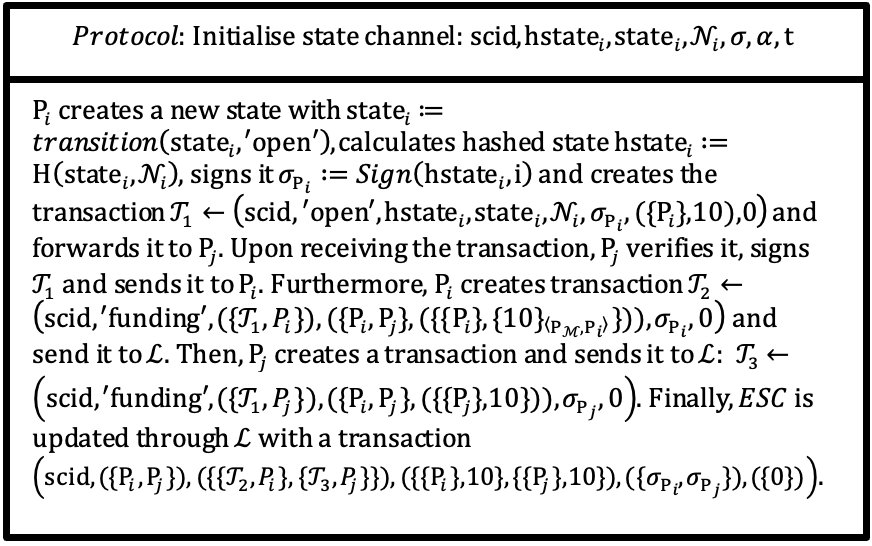}
\vspace*{-5mm}
\caption{Protocol function of a state channel initialisation.}
\label{fig:prot_init}
\end{figure}

\begin{figure}[htbp]
\centering
\includegraphics[width=\columnwidth]{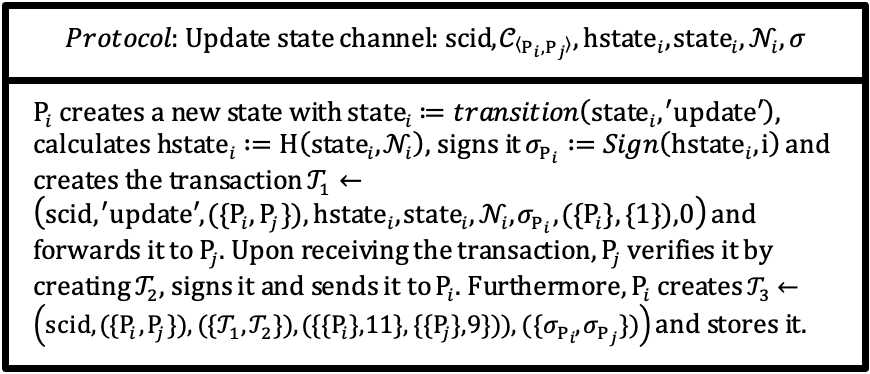}
\vspace*{-5mm}
\caption{Protocol function of a state channel update.}
\label{fig:prot_update}
\end{figure}

\begin{figure}[htbp]
\centering
\includegraphics[width=\columnwidth]{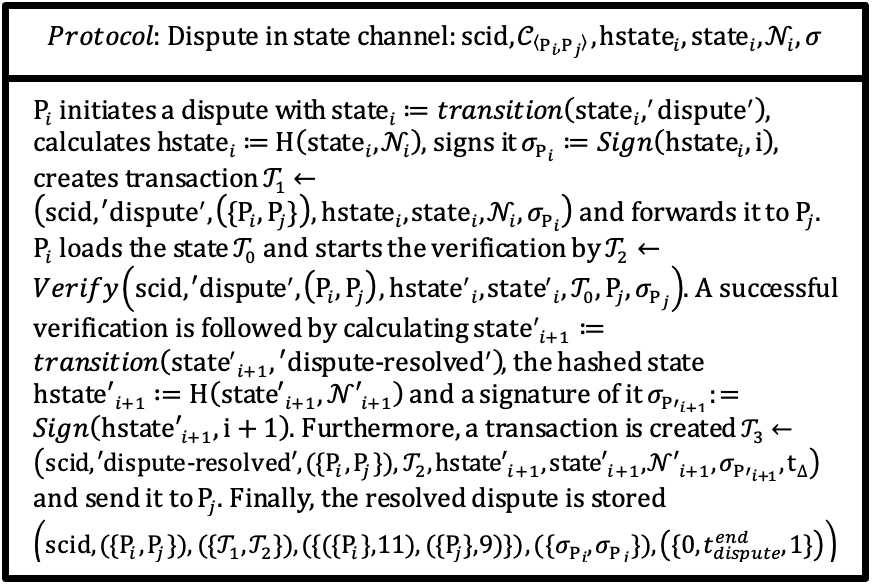}
\vspace*{-5mm}
\caption{Protocol function of a dispute in a state channel.}
\label{fig:prot_dispute}
\end{figure}

\begin{figure}[htbp]
\centering
\includegraphics[width=\columnwidth]{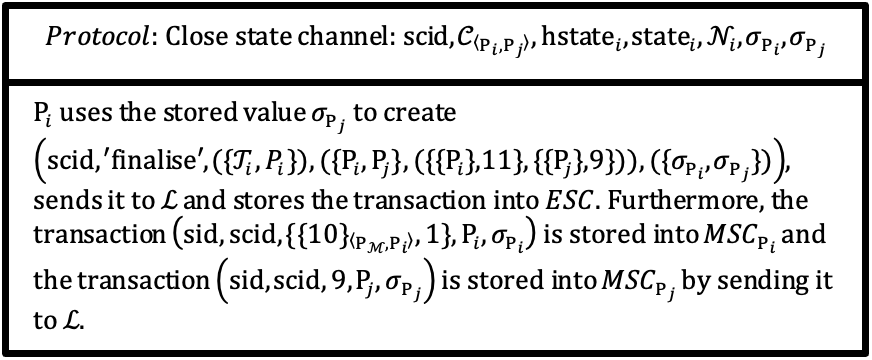}
\vspace*{-5mm}
\caption{Protocol function of a state channel closing.}
\label{fig:prot_close}
\end{figure}

Figures \labelcref{fig:prot_msc,fig:prot_esc,fig:prot_init,fig:prot_update,fig:prot_dispute,fig:prot_close} display our formalised protocol for the functionalities of $\mathcal{F}_{SMT}$, $\mathcal{F}_{SYN}$ and $\mathcal{L}$. Furthermore, the $id$ in the $MSC$ is assumed to be an identifier for the SCs, which are short for $id_i$ and can also be written as $(id, i)$. Additionally, we assume that the parties are ordered within a set $\mathcal{P}$. 
For better readability, we did not include any signature or verification keys into the $Sign$ or $Verify$ processes. We further assume that parties can compute transaction identifiers form the transactions $\mathcal{T}$ itself.

Figure \ref{fig:prot_msc} displays the protocol description of $MSC$ which starts with a party creating the new state comprising the current state and a command message. In the initialisation, the current state is not existent, which means that the state transition will become the command message $\text{state}_i := \text{`msc-init`}$. This state is then hashed together with a random nonce $\mathcal{N}$ and signed. After all information has been collected a transaction is created and send to the merchant. The merchant then verifies the transaction, creates an identifier, and returns it to the sender. 
If a participant wants to use the CNS (described in Section \ref{sec:creditnotesystem}), then collateral needs to be submitted to the merchant. However, the crucial part of this protocol is the next step where an additional key needs to be generated for the encryption of the collateral. For this purpose, the participant can use a similar generator function to the one described in Section \ref{sec:digitalsignaturescheme}. Instead of verification and signature keys, the generator function would create a public and a private key. Both keys (public and private key) are sent to the merchant in a secure channel but with the exception that the private key can only be accessed after a time window has elapsed. The time window's length can be determined based on the risk the merchant sees in the participant. However, the collateral is necessary to protect themselves from malicious behaviour. The time-lock mechanism, on the other hand, is necessary to protect the participant. The merchant then signs the collateral before creating the $MSC$.
Finally, the smart contract is created with a new session identifier $sid$, the identities of the merchant and the participant, the price $\alpha(\cdot)$, and signatures of both parties. 
The price within $MSC$ is defined as $\alpha^{\text{final}}_{\left\langle \text{P}_{\mathcal{M}},\text{P}_{i} \right\rangle}$ which comprises the total amount of coins of $\mathcal{M}$ $\alpha^{\text{final}}_{\text{P}_{\mathcal{M}}}$ and the total number of coins for $\text{P}_{i}$ $\alpha^{\text{final}}_{\text{P}_{i}}$.

Afterwards, the signed collateral can be used by a participant from the $ESC$. In Figure \ref{fig:prot_esc}, the description of the $ESC$ protocol is illustrated. As previously, a party completes the preparations before the initialisation transaction is created. 
For the $ESC$, each participant creates new keys through the generator function described in Section \ref{sec:digitalsignaturescheme}. These keys are then used for the whole communication within the $ESC$ and also for all communications in the state channels created in the future.
The other party then verifies the signature and creates a SC identifier $\text{scid} \gets \text{SC}(\left\langle P_i, P_j \right\rangle, n)$. The SC identifier $\text{scid}$ is created with the tuple of both parties' identities $\left\langle P_i, P_j \right\rangle$ and an incremental counter $n$ as inputs. Afterwards, $\text{scid}$ is send to the other participant. 
If any party uses the CNS, the other parties are notified subsequently regarding existing collateral. 
In practice, the participants might want to verify the signature of $\mathcal{M}$ on the collateral. We omitted this step for this version of the protocol but want to expand on this issue in a future paper. For this paper, we assume that the participants trust the signed collateral.
The following step is the creation of the $ESC$ and the update of corresponding $MSC$ contracts for both participants.

With an $ESC$, both parties can now initialise state channels $\mathcal{CH}$ as shown in Figure \ref{fig:prot_init}. Similarly to the previous protocols, one participant completes all preparations and sends a transaction to the other participant. This transaction comprises the data necessary to create the state channel. $\mathcal{CH}$'s initialisation message includes the identifier $scid$, command message, hashed state, state, nonce, signature, amount of coins certified with an identity, and a timer which can be used for the time-lock or the dispute timer.

Then, both participants need to fund the state channel. This can be accomplished by either sending their funds to the ledger $\mathcal{L}$ which locks the funds, or the CNS can be used in which, instead of the coins, only the signed collateral is sent to the ledger. 
In the case of the CNS, the participants need to ensure that the collateral is paid after the transactions are finished. This is a process that needs to be pursued by the merchant. 
Especially if the state channel is formed between two parties $P_i$ and $P_j$ where $P_i$ is using the CNS. Then $P_j$ needs to be reimbursed after the state channel closure by the merchant $\mathcal{M}$ as the merchant holds the collateral from the contract with $P_i$.  
After the funding is completed, the $ESC$ will get updated with the data of the current transactions. However, the number of coins in $ESC$ will only reflect the current amounts and not the final amount. This means that the collateral used by $P_i$ will not be reflected but instead shown as a locked coin amount.

The protocol function for a state channel update is illustrated in Figure \ref{fig:prot_update}. Here, a party proposes a state channel update by sending a command message $'update'$ with the signed state and a suggested new coin balance to the other party. Upon receiving that transaction, the signature, the new state and the updated balance need to be verified with the use of the $Verify(\cot)$ function. If the verification process was successful, the second party needs to sign the transaction and send it back. With the second signature, the transaction becomes valid and the channel update is completed. The new balance is now saved into the state channel.

For illustration purposes, we assume that a dispute is raised by $\text{P}_i$ (see Figure \ref{fig:prot_dispute}). Now, when a dispute is raised both participants have the opportunity to confirm whether the dispute was correctly appointed. For this purpose, the first step is to appoint the transaction which caused the dispute (i.e., sending the last transaction to  $\text{P}_j$). Then, the verification process is started through the $Verify(\cot)$ function by which the last transaction is checked and verified. The dispute can then end in two ways; either the dispute is resolved which means that the state channel can be continued or it is not resolved, in which case the state channel would be closed. Either way, the raised dispute will be remarked. 
Note that a dispute limit could be introduced which would punish too many raised disputes by listing a participant first on a grey list and later on a blacklist if enough disputes are raised. However, the participant would only be listed within the $ESC$ but the blacklist can also be expanded to the P2P marketplace.

Finally, Figure \ref{fig:prot_close} shows the process of closing the state channel. The closure of the state channel can be triggered in many ways. For instance, if the locked funds have been used up, a marketplace transaction is completed, a timer has elapsed, or a dispute was not resolved, which would trigger a channel closure.
The transactions until this point are processed and submitted to the ledger $\mathcal{L}$ by one party. Subsequently, the $'finalise'$ transaction is saved into $ESC$ is also processed to update the initial transactions which are written into both $MSC$ contracts. 
Especially, the coin amounts need to be finalised. For the party $\text{P}_i$, which used the CNS, the coin amount is not finalised directly but sent to the merchant which recalculates the final coin amount and also signs the recalculated coin amount. Furthermore, if the participant spends any coins during the transaction (i.e., the final coin amount would be less than what the current collateral displays), then the merchant is refunding the appropriate coins and sends them to the other party $\text{P}_j$. 
Note that we assume that the merchant has access to the payment address of the other party through $\text{P}_i$.

The protocol descriptions above are modelled after a best-case scenario. However, in case some error is created, the corresponding protocol that is in use needs to be aborted (as described in the ideal functionalities above). As such, we note that if a protocol aborts while a party creates any transaction, the party also aborts the protocol. Also, a party can be requested by any functionality to resend a state. If the state was already committed to a ledger, the party can send the committed transaction back. If the state was not committed, the transaction can be replicated but only if the transaction was not signed by every party. If it was signed and is therefore valid, the protocol needs to be either restarted (i.e., aborted and started again) or a dispute can be raised if malicious intend was ascertained.

\section{Evaluation}
\label{sec:evaluation}


In this section, we evaluate the security of the protocol based on Scyther and the UC framework.

\subsection{Protocol Evaluation}

For the protocol evaluation, we will use the formal analysis tool Scyther \cite{cremersScytherSemanticsVerification2006}. Scyther is a tool for formal verification that can verify the security requirements of cryptographic protocols \cite{cremersScytherSemanticsVerification2006, cremersScytherToolVerification2008}. For the security verification, Scyther uses the Dolev-Yao threat model \cite{dolevSecurityPublicKey1983} but the tool can support other threat models as well.
The tool can use functions, variables and constants to model the protocol using the Security Protocol Description Language (SPDL). In order to set the security requirements and objectives, a `claim` event is utilised in SPDL. 

For our protocol evaluation, we will set the objectives of secrecy of data, aliveness of a protocol, weak agreement, non-injective agreement (i.e., events happen in an expected sequence) and non-injective synchronisation (i.e., communicating parties agree with the exchanged content at the end of the protocol execution) \cite{cremersScytherSemanticsVerification2006, cremersScytherToolVerification2008}.
A protocol run ends with an output summary which provides feedback about the security and highlights if a potential attack is identified.

\begin{figure}[htbp]
\centering
\includegraphics[width=0.7\columnwidth]{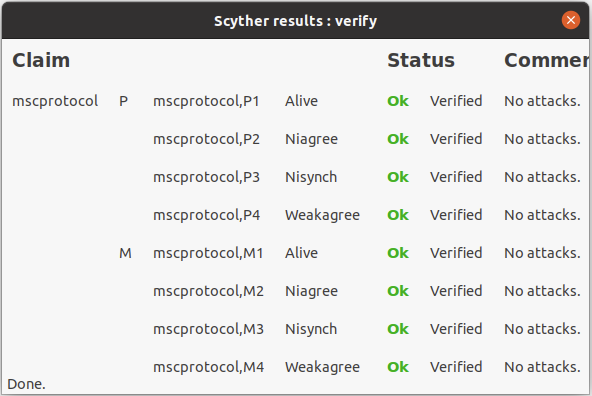}
\caption{MSC's security analysis results of Scyther.}
\label{fig:scyther_msc}
\end{figure}

\begin{figure}[htbp]
\centering
\includegraphics[width=0.7\columnwidth]{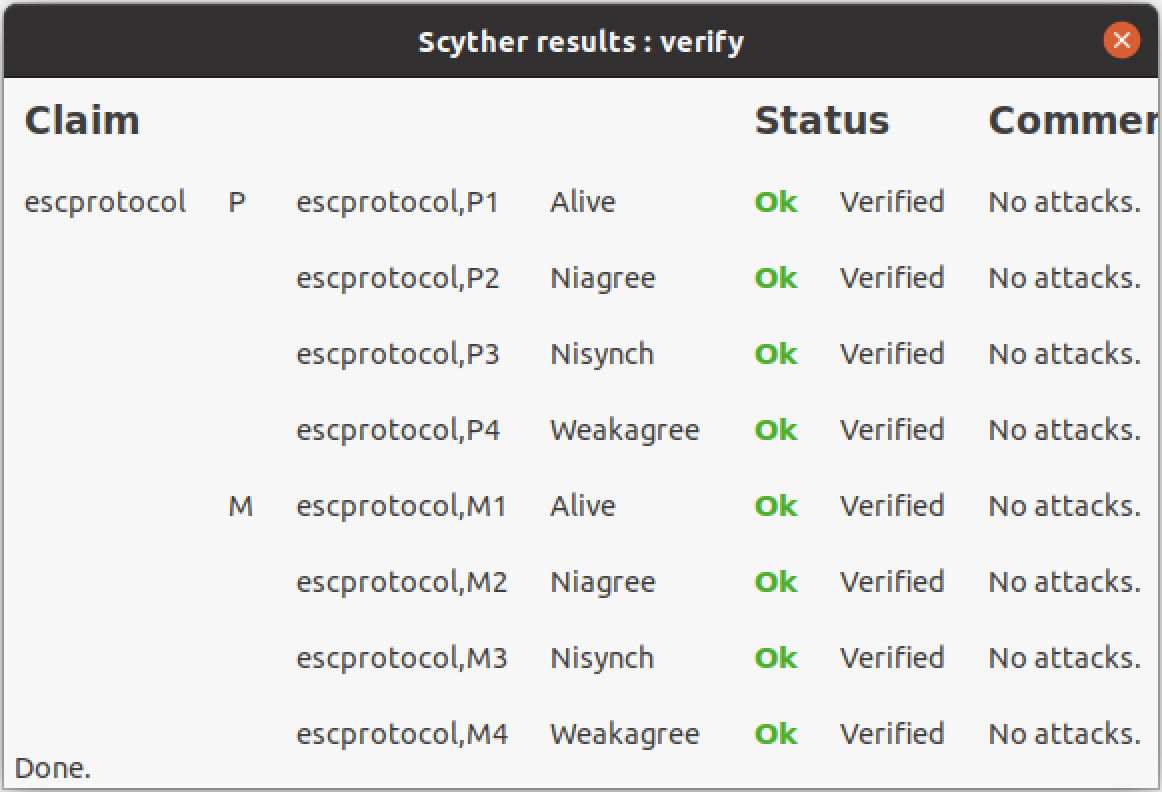}
\caption{ESC's security analysis results of Scyther.}
\label{fig:scyther_esc}
\end{figure}

\begin{figure}[htbp]
\centering
\includegraphics[width=0.85\columnwidth]{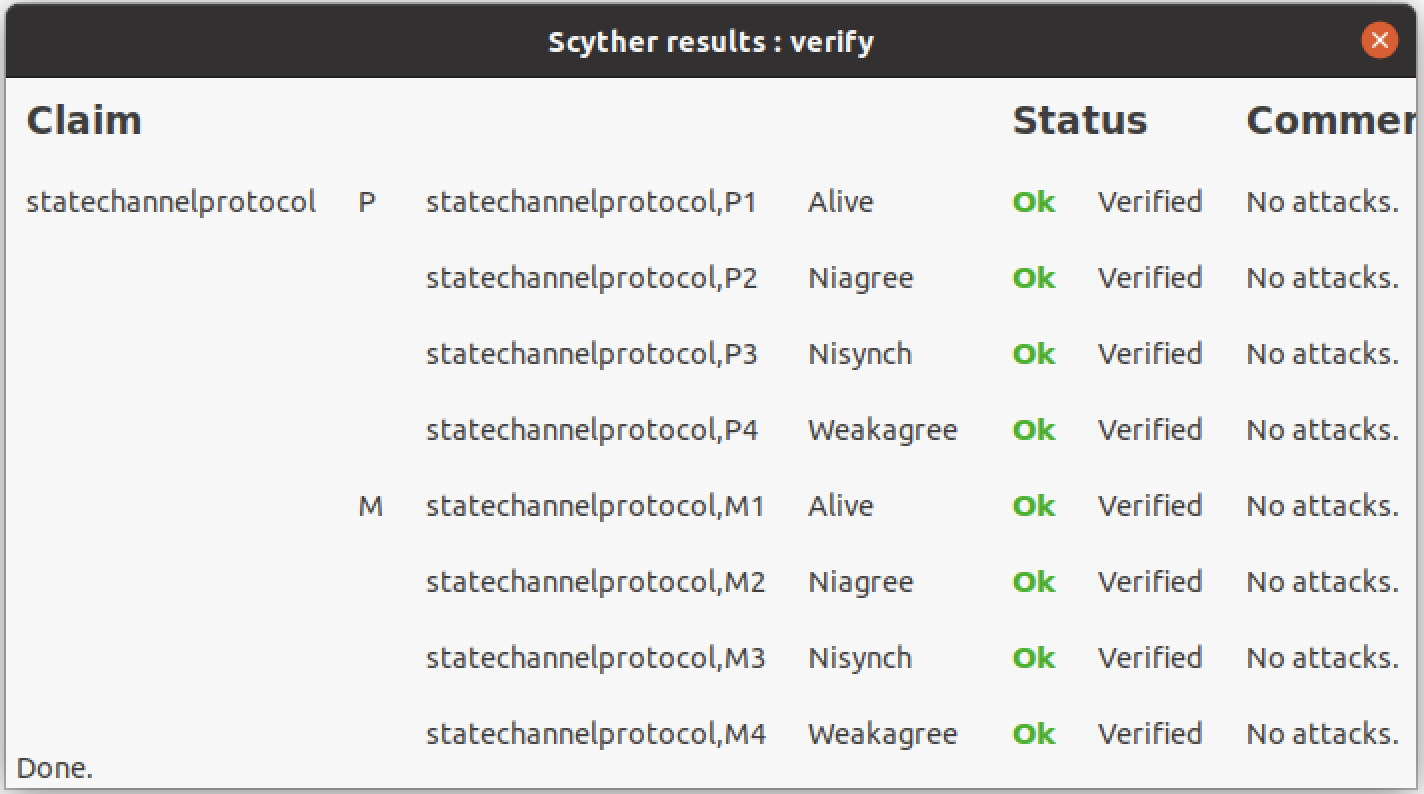}
\caption{State channel's security analysis results of Scyther.}
\label{fig:scyther_state_channel}
\end{figure}

For the formal verification, we modelled each of the protocol parts separately. The code is available at \url{https://github.com/jkalbantner/scyther_sc_protocol} \cite{kalbantnerScytherCodeDLTbased2021}. The results for the MSC (see Figure \ref{fig:scyther_msc}), ESC (see Figure \ref{fig:scyther_esc}) and for the state channel (see Figure \ref{fig:scyther_state_channel}) show that the Scyther tool could not find any attacks for the selected objectives.

Notice that Scyther has limitations when applied to our protocol as the tool was not designed for Blockchain systems. For instance, limitations apply to the simulation of specific blockchain security requirements such as atomicity. However, Scyther still does provide valuable security assurances under a Dolev-Yao adversary such as for confidentiality of any secret data, integrity, authenticity and replay protection.


\subsection{Security Analysis}

In Section \ref{sec:system_model}, we discussed how the ideal functionality $\mathcal{F}$ achieves our protocol goals (see Section \ref{sec:protocol_goals}). 
In particular, the security of our protocol is established in Theorem \ref{th:theorem_security} which shows that the protocol UC-realises $\mathcal{F}$ and therefore provides privacy, and atomicity. As the protocol can be used within most DLT systems, we assume that the used DLT system captures the properties of decentralisation, transparency, autonomy and non-repudiation respectively transaction immutability \cite{linSurveyBlockchainSecurity2017, zhengOverviewBlockchainTechnology2017}.

\begin{theorem}
\label{th:theorem_security}
Let the digital signature scheme from Section \ref{sec:digitalsignaturescheme} be EU-CMA (Existential Unforgeability under a Chosen Message Attack) secure \cite{goldwasserDigitalSignatureScheme1988}, then the protocol UC-realises the ideal functionality $\mathcal{F}$ in the model of ($\mathcal{F}_{SMT}, \mathcal{F}_{SYN}$).
\end{theorem}

\begin{IEEEproof}
For the verification of Theorem \ref{th:theorem_security}, let a simulator $\mathcal{S}$ interact with the ideal functionality $\mathcal{F}$ and emulate an indistinguishable protocol execution for adversary $\mathcal{A}$ against $\pi$. $\mathcal{S}$ forwards all corruption requests to $\mathcal{F}$ while maintaining a set of signing keys for honest parties so that $\mathcal{S}$ can sign transactions for them.
If $\mathcal{S}$ receives an invalid message from $\mathcal{A}$ then the execution will be aborted. 
\end{IEEEproof}

\paragraph{Privacy}
    For a protocol to be considered private, let the content of a transaction $\mathcal{T}$ be only accessible by a participant from set $\mathcal{P}$ within $MSC$ or $ESC$ executed in $\pi$. Furthermore, let the identities of $\mathcal{P}$ be only known to merchants from the set $\mathcal{M}$.
    The first condition is easily met as our protocol executes P2P only among protocol parties on secret and authenticated channels.
    However, the second condition requires the use of unique identifiers as proposed in Section \ref{sec:discussion}. In particular, the functionality needs to generate UUTID during a setup phase. This UUTID is then exchanged for their true identities. The parties' true identities will then be sent during the $MSC$ setup process to $\mathcal{M}$ which needs to store them separately. All commits to the ledger will contain the hidden identity. In order to further enhance privacy, a new identity could be generated for every $ESC$. In this case, however, the CNS could not be used as the collateral is still signed by the merchant and therefore could leak information about the identity of the user. 
    Alternatively, the hidden identity within the $MSC$ could be exchanged in a fixed time interval. This ruling could also apply to the regeneration of signatures and verification keys. 
    
\paragraph{Atomicity}
    As for the property of atomicity, let the output to every transaction $\mathcal{T}$ be either one or zero (i.e., executed completely or not at all). 
    
    For every transaction that $\mathcal{S}$ receives from $\mathcal{F}$, a response is sent on behalf of an honest user to $\mathcal{A}$. When $\mathcal{A}$ rejects, it sends a message including $\bot$ to the functionality and aborts the simulation, otherwise it sends $\top$.
    $\mathcal{S}$ assumes that honest parties and compromised parties (controlled by $\mathcal{A}$) can receive messages. 
    For instance, if $\mathcal{S}$ awaits a message and $\mathcal{A}$ does not send responses for all compromised users, $\mathcal{S}$ will respond with $\bot$ and will abort; otherwise, $\mathcal{S}$ will send $\top$.
    Additionally, $\mathcal{S}$ can simulate the creation of signatures for honest users and send them to $\mathcal{A}$. If $\mathcal{A}$ produces signatures for all corrupted parties, $\mathcal{S}$ sends $\top$ and otherwise $\bot$.
    
    As for disputes, $\mathcal{F}$ creates a dispute for the last transaction and sends it to $\mathcal{S}$.
    $\mathcal{S}$ then waits for the verification from $\mathcal{F}$. Without a response, $\mathcal{S}$ will send $\bot$. Otherwise, $\mathcal{S}$ simulates the verification of the signature for the honest parties and waits for the dispute-resolved transaction within the dispute timer's time window $\text{t}^{now} < \text{t}^{end}_{dispute}$. All transactions are also sent to $\mathcal{A}$. Upon receiving the transaction, $\mathcal{S}$ will respond with $\top$; otherwise, it is considered a fail and it sends $\bot$.
    For the same process, if $\mathcal{A}$ does not send a verification response to $\mathcal{S}$ within the time window, the process will be aborted.

\section{Application}
\label{sec:application}

The described protocol and model can be applied to many scenarios. However, we choose to focus on a specific use case of P2P marketplaces. 
A P2P marketplace is a decentralised electronic marketplace that inherits the advantages of distributed architectures (e.g., fail-safety, lower costs, scalability, performance) \cite{ferreiraBuildingEmarketplacePeertopeer2004}. 
DLT systems themselves are constructed for decentralised P2P communication which makes them predestined for the use within P2P marketplaces \cite{laoSurveyIoTApplications2020,linSurveyBlockchainSecurity2017}. There are various specialisations for P2P marketplace but we will concentrate on the application of them within smart grids. In particular, the application of our protocol to P2P marketplaces that trade energy contracts as described by Kalbantner et al. \cite{kalbantnerP2PEdgeDecentralisedScalable2021}. The authors \cite{kalbantnerP2PEdgeDecentralisedScalable2021} not only propose the use of DLT architecture but also include state channels to improve scalability. However, their model is focusing on the marketplace and the trading on the marketplace rather than on the state channel; hence, our protocol fills a gap in their model.

\subsection{Use Case: A Study Of Grid Transactions}

Smart grids can be described as decentralised systems with a bi-directional data and energy flow between customers and producers \cite{sianoSurveyEvaluationPotentials2019,andoniBlockchainTechnologyEnergy2019,armbrustViewCloudComputing2010, kalbantnerP2PEdgeDecentralisedScalable2021}. 
Compared to their traditional counterparts, the power grid, smart grids can provide better scalability, performance, and resilience. 
Numerous researchers \cite{zhangEnergyTradingDemand2019,liConsortiumBlockchainSecure2017,gaiPermissionedBlockchainEdge2019,luoDistributedElectricityTrading2019, kalbantnerP2PEdgeDecentralisedScalable2021} have proposed decentralised approaches to implement smart grids' control, business and management processes based on DLT systems. The model by Kalbantner et al. \cite{kalbantnerP2PEdgeDecentralisedScalable2021} describes a multi-layered electricity market model which comprises (i) a DLT marketplace, (ii) an off-chain state channel for any energy contract negotiation, (iii) a communication layer for their model and (iv) the electrical grid itself. We refer to the paper itself for a more detailed view of the model. 

Energy marketplaces comprise multiple stakeholders that could be interested in participating in the marketplace. Firstly, there are consumers which are marketplace participants that can only buy energy. Secondly, prosumers are participants that can trade energy contracts (i.e., buy energy) but can also generate energy and also sell energy to the marketplace. Thirdly, there are producers which are parties that can only sell energy and do not buy any energy from the marketplace. Finally, there are distributors otherwise known as Distribution System Operators (DSO) which are intermediaries that own or rent part of the power grid infrastructure. These participants usually provide part of the energy distribution infrastructure \cite{kalbantnerP2PEdgeDecentralisedScalable2021}.

\subsection{Modifications}

In the current version, the protocol description is flexible so that it can be applied to a multitude of scenarios. 
To use our protocol in an energy marketplace model \cite{kalbantnerP2PEdgeDecentralisedScalable2021}, however, some adjustments are necessary. Notice that every alteration of the protocol must not change the protocol goals as defined in Section \ref{sec:protocol_goals}.

\textit{$MSC$}.
    The $MSC$ (displayed in Figure \ref{fig:func_msc}) needs to be adjusted to be able to trade energy contracts for any number of coins. Furthermore, the merchant $\mathcal{M}$ of our protocol refers to a participant who can sell energy. This could be, for instance, a Transmission System Operator (TSO) which is the operator of the power grid or a distributor. 
    
    In order to adjust the protocol, we need to include the energy contracts. In particular, Kalbantner et al. \cite{kalbantnerP2PEdgeDecentralisedScalable2021} included the identity of the buyer or the seller, the price of an energy contract, the quantity for the trade and a time variable into their smart contracts. 
        
    Most of the elements are already available as $MSC_{\text{P}_i}$ is defined as follows (see Figure \ref{fig:func_msc}):

    \begin{equation}
         ( \text{sid}, \mathfrak{S} , (\{\mathcal{M},\text{P}_i\}) , \alpha^{final}_{\left\langle \text{P}_\mathcal{M} , \text{P}_i \right\rangle} , 
        (\{ \alpha_{\mathcal{M}} , \alpha_{\text{P}_i} \}) )
    \end{equation}
    
    Thus, we include the quantity of energy $\mathcal{Q}$ into the $MSC$ which results in the following contract:
    \begin{multline}
        ( \text{sid}, \mathfrak{S} , (\{\mathcal{M},\text{P}_i\}) , (\{ ( \alpha^{final}_{\text{P}_i } , \mathcal{Q}^{final}_{\text{P}_i} ) , \\ (\alpha^{final}_{\text{P}_j } , \mathcal{Q}^{final}_{\text{P}_j}) \}) , (\{ \alpha_{\mathcal{M}} , \alpha_{\text{P}_i} \}) )
    \end{multline}
    
    The quantity of energy $\mathcal{Q}^{final}_{\left\langle \text{P}_\mathcal{M} , \text{P}_i \right\rangle}$ is the finalised energy amount which was returned from the $ESC$ after the negotiation ended. 
    Notice that the quantity $\mathcal{Q}$ is sold in energy packets.

\textit{$ESC$}.
    Similarly to the $MSC$, the $ESC$ (illustrated in Figure \ref{fig:func_esc}) is currently not able to contain energy contracts. Currently, $ESC$ is defined as follows:
    \begin{equation}
        ( \text{scid}, \mathbb{C}_{\left\langle \text{P}_i , \text{P}_j \right\rangle} , \mathbb{T}_{\left\langle \text{P}_i , \text{P}_j \right\rangle} , 
        \alpha_{\left\langle \text{P}_i , \text{P}_j \right\rangle} , \alpha^{final}_{\left\langle \text{P}_i , \text{P}_j \right\rangle} , \mathbb{D}_{\left\langle \text{P}_i , \text{P}_j \right\rangle} )
    \end{equation}
    
    With $\mathcal{Q}_{\left\langle \text{P}_i , \text{P}_j \right\rangle}$ as the temporary contractual energy quantities for $\text{P}_i$ and $\text{P}_j$ and the final contractual energy quantities $\mathcal{Q}^{final}_{\left\langle \text{P}_i , \text{P}_j \right\rangle}$ for $\text{P}_i$ and $\text{P}_j$ the $ESC$ needs to be modified to the following:
    \begin{multline}
        ( \text{scid}, \mathbb{C}_{\left\langle \text{P}_i , \text{P}_j \right\rangle} , \mathbb{T}_{\left\langle \text{P}_i , \text{P}_j \right\rangle} , (\{ ( \alpha_{\text{P}_i } ,  \mathcal{Q}_{\text{P}_i} ) , (\alpha_{\text{P}_j } , \mathcal{Q}_{\text{P}_j}) \}) , \\
        (\{ ( \alpha^{final}_{\text{P}_i } , \mathcal{Q}^{final}_{\text{P}_i} ) , (\alpha^{final}_{\text{P}_j } , \mathcal{Q}^{final}_{\text{P}_j}) \}) , \mathbb{D}_{\left\langle \text{P}_i , \text{P}_j \right\rangle} )
    \end{multline}
    
    Additionally, the update of of the $MSC$ contracts need to be altered to include the quantity. The modified version for participant $\text{P}_i$ is as follows (symmetrical for $\text{P}_j$):
    \begin{equation}
        \left( \text{sid}, \text{scid}, \text{P}_i , (\{ ( \alpha_{\text{P}_i } , \mathcal{Q}_{\text{P}_i} ) , (\alpha_{\text{P}_j } , \mathcal{Q}_{\text{P}_j}) \}) , \sigma_{\text{P}_i} \right)
    \end{equation}

\textit{State channels}.
    Currently, the state channel (Figures \labelcref{fig:func_init,fig:func_update,fig:func_dispute,fig:func_close}) describe only the negotiation and exchange of coin amounts. 
    The marketplace model from \cite{kalbantnerP2PEdgeDecentralisedScalable2021} is orchestrating the contract negotiation process. In particular, the state channels' task is to conduct the negotiation which means that the state channels must be adjusted to allow quantities of energy contracts and prices to be exchanged.
    However, as the state channel stays the same throughout every step (i.e., initialisation, update, dispute, closure), we will concentrate on the update functionality of the state channel (see Figure \ref{fig:func_update}). 
    The current process from the perspective of party $\text{P}_i$ is as follows (illustrated in Figure \ref{fig:func_update}):
    \begin{multline}
        ( \text{scid},\text{'update'},\mathcal{C}_{\left\langle \text{P}_i , \text{P}_j \right\rangle},\text{hstate}_i,\text{state}_i,\mathcal{N}_i, \\
        \alpha_{\text{P}_i},\sigma_{\text{P}_i},\text{t}_\Delta )
    \end{multline}
    
    With the addition of the quantity $\mathcal{Q}_{\text{P}_i}$ the process is changed as follows:
    \begin{multline}
        ( \text{scid},\text{'update'},\mathcal{C}_{\left\langle \text{P}_i , \text{P}_j \right\rangle},\text{hstate}_i,\text{state}_i,\mathcal{N}_i, \\
        (\{ ( \alpha_{\text{P}_i } , \mathcal{Q}_{\text{P}_i} ) \}),\sigma_{\text{P}_i},\text{t}_\Delta )
    \end{multline}
    
    After a successful verification by the other party (i.e., $\text{P}_j$), an update will be send by $\text{P}_i$ which is modified as follows:
    \begin{multline}
        ( \text{scid},\text{'update-success'},\mathcal{C}_{\left\langle \text{P}_i , \text{P}_j \right\rangle}, \\
        (\{ ( \alpha_{\text{P}_i } , \mathcal{Q}_{\text{P}_i} ) ,  (\alpha_{\text{P}_j } , \mathcal{Q}_{\text{P}_j}) \}), 
        (\{\sigma_{\text{P}_i} , \sigma_{\text{P}_i} \}),\text{t}_\Delta )
    \end{multline}
    
    Upon receiving the message, $\text{P}_j$ stores the finalised amounts (e.g., $\alpha^{final}_{\text{P}_i }$ and quantities $\mathcal{Q}^{final}_{\text{P}_i }$):
    \begin{multline}
        ( 
        \text{scid},
        \mathcal{C}_{\left\langle \text{P}_i , \text{P}_j \right\rangle},
        (\{ \mathcal{T}_i, \mathcal{T}_{i+1} \}),
        (\{ ( \alpha^{final}_{\text{P}_i } , \mathcal{Q}^{final}_{\text{P}_i} ) , \\ (\alpha^{final}_{\text{P}_j } , \mathcal{Q}^{final}_{\text{P}_j}) \}), 
        \sigma_{\text{P}_{i+1}}
        )
    \end{multline}

\section{Conclusion}
\label{sec:conclusion}

In this work, we present a protocol that combines a hierarchical smart contract infrastructure with state channels to increase scalability of an underlying DLT system. 
The presented protocol proposes a novel approach to optimise collateral usage in state channels.  
Accordingly, we propose the credit-note system (CNS) which can be used by protocol participants to trade with collateral which can save time, reduce management overhead and lower the costs for involved parties. We present the goals of the protocol in Section \ref{sec:protocol_goals} and evaluate the protocol's security in Section \ref{sec:evaluation}.

Furthermore, in Section \ref{sec:application} we consider the protocol in the context of smart grid marketplaces that trade energy contracts. We evaluate a use case in which there is constant, unpredictable demand on the system and incentive for attackers to abuse the underlying technology for both financial and political reasons. Our protocol can help in these environments by providing robust, timely dispute resolution mechanisms which leverage the mutual stake in the contracts being executed to ensure that non-compliance is a costly endeavour. Subsequently, the mechanisms also need to ensure that false positives are trivial to remediate based on a cursory analysis of blocks that have been submitted between the initial allegation of bad faith and the subsequent discovery of a false accusation.

As future work, we intend to expand on the usage of collateral in future DLT applications as well as expand on the verification of participant's collateral through zero-knowledge proofs. Additionally, we want to expand our state channels so that an $n-m$ relationship between multiple participants can be realised and that there are no restrictions on the number of participants. Finally, we want to enhance the credit-note system so that credit-notes and collateral can be changed after a state channel initialisation. Additionally, participants should be able to fund state channels while they are active which would make them more efficient.

\ifCLASSOPTIONcompsoc
\else
\fi


\ifCLASSOPTIONcaptionsoff
  \newpage
\fi


\bibliographystyle{IEEEtran}
\bibliography{IEEEabrv,bibliography/MyLibrary}
%



%


\begin{IEEEbiography}[{\includegraphics[width=1in,height=1.25in,clip,keepaspectratio]{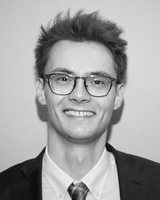}}]{Jan Kalbantner}
received the BS in computer science from Cooperative State University BW, Mosbach, Germany and the MS in information security from Royal Holloway, University of London, UK, in 2016 and 2019, respectively. Since 2019, he is working towards the PhD degree at Royal Holloway, University of London, UK. His research interests include cyber physical systems, DLT, network security, and secure architectures.
\end{IEEEbiography}


\begin{IEEEbiography}[{\includegraphics[width=1in,height=1.25in,clip,keepaspectratio]{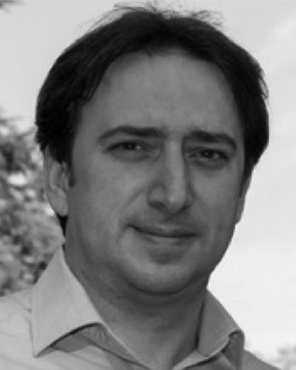}}]{Konstantinos Markantonakis}
received the BS in computer science from Lancaster University, Lancaster, UK, in 1995, the MS in information security from Royal Holloway, University of London, London, U.K., in 1996, the PhD in information security from Royal Holloway, University of London, London, U.K., in 2000, and the MBA in international management from School of Management, Royal Holloway, University of London, London, UK, in 2005. He is currently the Director of the Smart Card and IoT Security Centre. He has authored or coauthored more than 190 papers in international conferences and journals. His main research interests include smart card security and applications, embedded system security and trusted execution environments, cyber physical systems, and Internet of Things.
\end{IEEEbiography}


\begin{IEEEbiography}[{\includegraphics[width=1in,height=1.25in,clip,keepaspectratio]{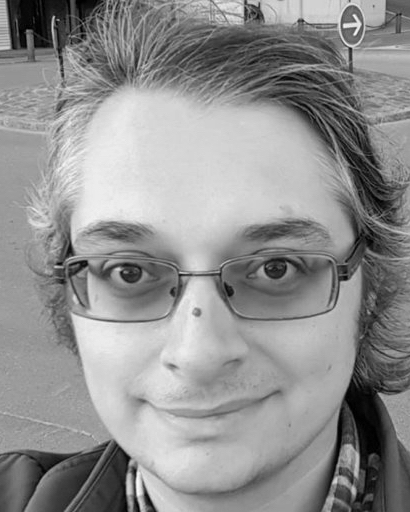}}]{Darren Hurley-Smith} 
received the BEng in computer systems and hardware design and the PhD in computer and network security from University of Greenwich in 2012 and 2015, respectively. He is currently a Lecturer in information security at the Information Security Group at Royal Holloway, University of London, UK. His interests are in statistical testing of random number generators, RFID/NFC security, mobile Ad Hoc network security, and cryptocurrency. He also has a keen interest in ransomware, the economics of cyber-crime, and autonomous aerial systems.
\end{IEEEbiography}


\begin{IEEEbiography}[{\includegraphics[width=1in,height=1.25in,clip,keepaspectratio]{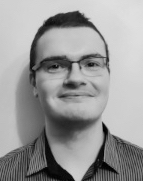}}]{Carlton Shepherd}
received his PhD in information security from Royal Holloway, University of London, UK, and his BS in computer science from Newcastle University, UK. He is currently a Senior Research Fellow at the Information Security Group at Royal Holloway, University of London, UK. His research interests centre around the security of trusted execution environments (TEEs) and their applications, secure CPU design, embedded systems, applied cryptography, and hardware security.
\end{IEEEbiography}


\begin{IEEEbiography}[{\includegraphics[width=1in,height=1.25in,clip,keepaspectratio]{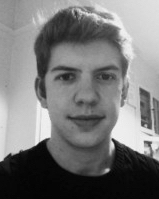}}]{Benjamin Semal}
received the MEng in electrical engineering from Ecole Polytechnique Universitaire of Montpellier and the MS in robotics from Cranfield University in 2016. In 2017, he worked as a hardware security analyst at UL Transaction Security, where he performed fault attacks on secure processors for EMVCo certification. In 2018, he joined Royal Holloway University of London to work towards the PhD in information security. His research interest focused mainly on microarchitectural attacks for information leakage in multi-tenant environments. Since 2020, Benjamin works as a security engineer at SERMA SECURITY \& SAFETY, where he conducts evaluation of point of sale devices and cryptographic modules.
\end{IEEEbiography}








\end{document}